\documentclass{jfm}
\usepackage{lineno,hyperref}
\usepackage{hyperref}
\usepackage{fancyhdr}
\usepackage{amsmath}
\usepackage{mathtools}
\usepackage{amsfonts}
\usepackage{amssymb}
\usepackage{textcomp}
\usepackage{booktabs}                      
\usepackage{multicol}
\usepackage{multirow}
\usepackage{xspace}
\usepackage{nomencl}
\usepackage{subcaption,booktabs}
\usepackage{caption}
\usepackage{float}
\usepackage[dvipsnames]{xcolor}
\usepackage{rotating}
\makenomenclature
\usepackage{tabularx}                      
\usepackage[square,sort,comma,numbers]{natbib}
\usepackage{adjustbox}
\usepackage[graphicx]{realboxes}
\usepackage[toc,page]{appendix}
\usepackage{cancel}
\usepackage{graphicx}
\usepackage{subcaption}
\usepackage{epstopdf}
\usepackage{color}

\let\OldS\S
\renewcommand{\S}{\OldS\xspace}

\newcommand{\nsc}[1]{{\color{black}#1}}

\shorttitle{Evaporating Rayleigh-B\'enard convection}

\shortauthor{N. Scapin, A. D. Demou, L. Brandt}

\title{Evaporating Rayleigh-B\'enard convection: prediction of interface temperature and global heat transfer modulation}
\author{
  Nicol\`o Scapin\aff{1}\corresp{\email{nicolos@mech.kth.se}},
  Andreas D. Demou\aff{2},
  Luca Brandt\aff{1,3}  
}
\affiliation
{
\aff{1}FLOW, Department of Engineering Mechanics, Royal Institute of Technology (KTH), Stockholm, Sweden,
\aff{2}Computation-based Science and Technology Research Center, \\ The Cyprus Institute, Nicosia, Cyprus,
\aff{3}Department of Energy and Process Engineering, Norwegian University of Science and Technology (NTNU), Trondheim, Norway.
}

\begin{document}

\maketitle
%
%
\begin{abstract} 
\nsc{
We propose an analytical model to estimate the interface temperature $\Theta_{\Gamma}$ and the Nusselt number $Nu$ for an evaporating two-layer Rayleigh-B\'enard configuration in statistically stationary conditions. The model is based on three assumptions: (i) the Oberbeck-Boussinesq approximation can be applied to the liquid phase, while the gas thermophysical properties are generic functions of thermodynamic pressure, local temperature, and vapour composition, (ii) the Grossmann-Lohse theory for thermal convection can be applied to the liquid and gas layers separately, (iii) the vapour content in the gas can be taken as the mean value at the gas-liquid interface. We validate this setting using direct numerical simulations (DNS) in a parameter space composed of the Rayleigh number ($10^6\leq Ra\leq 10^8$) and the temperature differential ($0.05\leq\varepsilon\leq 0.20$), which modulates the variation of state variables in the gas layer. To better disentangle the variable property effects on $\Theta_\Gamma$ and $Nu$, simulations are performed in two conditions. First, we consider the case of uniform gas properties except for the gas density and gas-liquid diffusion coefficient. Second, we include the variation of specific heat capacity, dynamic viscosity, and thermal conductivity using realistic equations of state. Irrespective of the employed setting, the proposed model agrees very well with the numerical simulations over the entire range of $Ra-\varepsilon$ investigated.
}
\end{abstract}
\begin{keywords}
Thermal convection, Evaporation, Non-Oberbeck–Boussinesq effects.
\end{keywords}
%
%
\section{Introduction}\label{sec:intro}
Evaporation on a horizontal gas-liquid interface plays a pivotal role in different contexts, from geophysical processes such as moisture convection and vapour distribution within the atmosphere~\citep{colman2021water}, to industrial applications such as the cooling of fuel rods in the spent-fuel pools of nuclear reactors~\citep{hay2020evaporation}. The presence of vapour in the gas changes the local thermophysical properties, in particular, the density and heat capacity~\citep{colman2021water}, and thus modifies the global heat transfer in the system, quantified by the Nusselt number, $Nu$~\citep{schumacher2010buoyancy}. The mean vapour content in the gas phase depends on its value at the interface, which in turn is a function of the partial pressure and the interface temperature in an exponential fashion (e.g.\ Clausius-Clapeyron law, Span-Wagner relation). Thus, small changes in the interface temperature significantly impact the amount of vapour in the gas and the total heat transfer. A conceptually simple set-up to study these flows in a precise and controlled manner is the multiphase Rayleigh-B\'enard (RB) configuration: two infinitely extended fluid layers confined by two horizontal walls at a fixed temperature, heated from below and cooled from above. Inspired from the classical single-phase counterpart used to model turbulent convection~\citep{ahlers2009heat,chilla2012new}, this configuration has been the object of numerical~\citep{nataf1988responsible,prakash1994convection_i} and experimental~\citep{xie2013dynamics,zhang2019moisture} studies. In particular, the multiphase RB set-up has been recently adopted to study: (i) the interface break-up in the presence of buoyancy~\citep{liu2021two}, (ii) the heat transfer enhancement due to the manipulation of the wall wettability~\citep{liu2022enhancing}, and (iii) the modulation of heat transfer and interface temperature induced by the variation of the liquid layer height and thermal conductivity of the two phases~\citep{liu2021heat}. All these numerical studies did not consider phase change and assumed constant thermophysical properties within the Oberbeck-Boussinesq (OB) approximation, with the exception of~\cite{biferale2012convection} for boiling flows and of~\cite{favier2019rayleigh} for ice melting.
\nsc{
Here, we include evaporation at the two-phase interface and relax the assumption of constant and uniform thermophysical properties in the gas phase while keeping the ones of the liquid uniform and constant. In particular, we extend the theory for the interface temperature proposed in~\cite{liu2021heat} to account for i) phase change at the interface, ii) non-Oberbeck-Boussinesq (NOB) effects in the gas phase induced by variations of the thermodynamic pressure, temperature, and composition. We propose analytical scaling laws for predicting the interface temperature and the heat transfer modulation with respect to a RB system without evaporation. The resulting expressions are compared against high-fidelity DNS, which are performed using a weakly compressible multiphase formulation with phase change, covering a substantial region of the $Ra-\varepsilon$ parameter space.
} \par
%
%
This paper is organized as follows. In \S~\ref{sec:scaling}, we introduce the main assumptions and derive analytical expressions for the interface temperature and heat transfer modulation in the evaporating Rayleigh-B\'enard system. In \S~\ref{sec:math_num}, we describe the mathematical and numerical model employed to validate the analytical scaling laws. In \S~\ref{sec:assess}, we present the validation of the theory and an assessment of the assumptions behind the model. The main findings and conclusions are summarized in \S~\ref{sec:concl}.
%
%
\section{Interface temperature and global heat transfer modulation}\label{sec:scaling}
\begin{figure*}
  \centering
  \includegraphics[width=12.32 cm, height=5.25 cm]{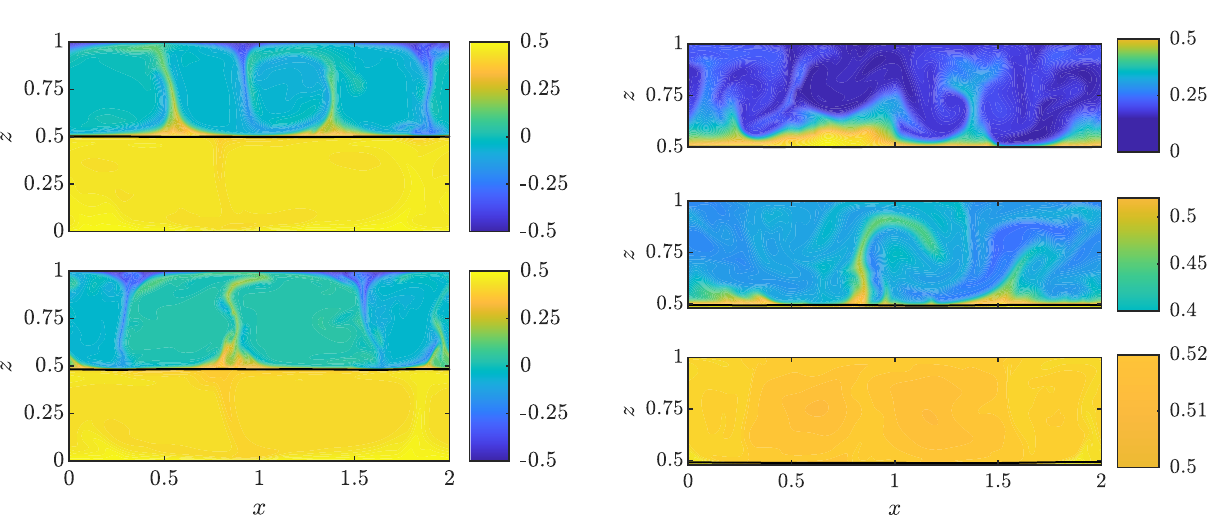}
  \put(-350,130){\small(\textit{a})}
  \put(-350,065){\small(\textit{b})}
  \put(-173,130){\small(\textit{c})}
  \put(-173,085){\small(\textit{d})}
  \put(-173,040){\small(\textit{e})}
  \caption{Multiphase Rayleigh-B\'enard convection for $Ra=10^8$ and $\varepsilon=0.20$ (taken from the numerical simulations). Left: snapshots of the temperature distribution $\Theta$ (\textit{a}) at the start of evaporation ($t=0$), and (\textit{b}) after a statistically stationary state is achieved ($t>300$). Right: snapshots of the vapour mass fraction, $Y_l^v$, distribution in the gas region at (\textit{c}) $t\approx 60$ (\textit{d}) $t\approx 122$ and (\textit{e}) steady state, $t>300$ ($t$ is scaled with the free-fall time $\widehat{t}_{ff}=\widehat{l}_z//(2\varepsilon|\widehat{\mathbf{g}}|\widehat{l}_z)^{0.5}$).}
  \label{fig:sca_vol}
\end{figure*}
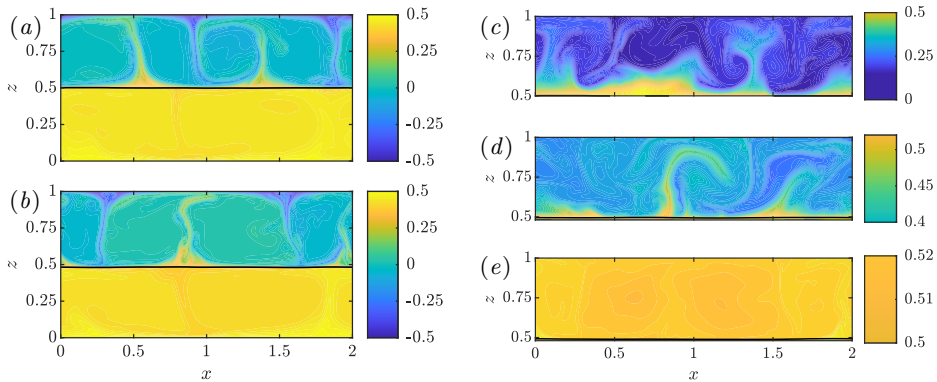

We consider a cavity partially filled with an evaporating single-component liquid and an initially dry gas, as shown in figure~\ref{fig:sca_vol}. The domain is laterally unbounded and confined by two horizontal walls separated by a distance $\widehat{l}_z$ ($\widehat{\cdot}$ indicates a dimensional quantity). Constant temperatures, $\widehat{T}_b=(1+\varepsilon)\widehat{T}_r$ and $\widehat{T}_t=(1-\varepsilon)\widehat{T}_r$ are imposed on the top heated and bottom cooled walls, where $\varepsilon=(\widehat{T}_b-\widehat{T}_t)/(2\widehat{T}_r)= \widehat{\Delta T}/(2\widehat{T}_r)$ is the dimensionless temperature differential and $\widehat{T}_r=(\widehat{T}_t+\widehat{T}_b)/2$ is the mean temperature, taken, hereinafter, as reference value. Under these conditions, the system eventually reaches a statistically stationary condition, with the vapour saturating the gas layer, leading to a dynamic balance between evaporation and condensation at the interface $\Gamma$. The presence of vapour dramatically changes the statistically stationary state: it modifies the gas thermophysical properties and hence the heat transfer inside the cavity, while reducing the height of the liquid layer. Here, we characterize these variations as a function of the amount of vapour inside the cavity. To this purpose, we write an expression for the mean interfacial vapour mass concentration $\overline{Y}_{l,\Gamma}^v$ (deduced from Raoult's law) and the Span-Wagner model for the interfacial vapour pressure $p_{s,\Gamma}$,
%
%
%
%
\begin{equation}
  \begin{cases}
      \overline{Y}_{l,\Gamma}^v = \dfrac{\lambda_Mp_{s,\Gamma}}{\lambda_Mp_{s,\Gamma}+(p_{th}-p_{s,\Gamma})}\mathrm{,} \\
      p_{s,\Gamma} = \Pi_P^{-1}\exp\left[B_1\eta_{sw}+B_2\eta_{sw}^{1.5}+B_3\eta_{sw}^{2.5}+B_4\eta_{sw}^{5})(1-\eta_{sw})^{-1}\right]\mathrm{.}\label{eqn:rault_sw_par}
  \end{cases}
\end{equation}
In equation~\eqref{eqn:rault_sw_par}, $\lambda_M=\widehat{M}_l/\widehat{M}_{g,r}$ is the molar mass ratio between liquid and inert gas and $\Pi_P=\widehat{p}_{th,r}/\widehat{p}_{cr}$ with $\widehat{p}_{th,r}$ the reference thermodynamic pressure and $\widehat{p}_{cr}$ the critical pressure. The quantity $\eta_{sw}=1-\widehat{T}_{\Gamma}/\widehat{T}_{cr}$ is the Span-Wagner parameter, and the coefficients $B_{i=1,4}$ depend on the substance under consideration. By introducing the dimensionless interface temperature $\Theta_{\Gamma} = (\widehat{T}_{\Gamma}-\widehat{T}_r)/\widehat{\Delta T}$, $\eta_{sw}$ becomes
\begin{equation} 
  \eta_{sw} = 1-\dfrac{\widehat{T}_{\Gamma}}{\widehat{T}_{cr}}=1-\Pi_T\left(1+2\varepsilon\Theta_{\Gamma}\right)\mathrm{,}
  \label{eqn:eta_sw}
\end{equation}
where $\Pi_T=\widehat{T}_r/\widehat{T}_{cr}$. Equations~\eqref{eqn:rault_sw_par} and~\eqref{eqn:eta_sw} show that fixing the type of substance, $\overline{Y}_{l,\Gamma}^v$ is determined by five quantities: (i) the ratio between the mean temperature and the critical temperature, $\Pi_T$, (ii) the ratio between the reference thermodynamic pressure and the critical pressure, $\Pi_P$, (iii) the temperature differential $\varepsilon$, (iv) the interface temperature $\Theta_\Gamma$, (v) the thermodynamic pressure $p_{th}$. The first three quantities depend on the ambient conditions (typically given or measured) and the type of substance, whereas $\Theta_{\Gamma}$ and $p_{th}$ depend on the flow in the two phases and is determined below. \par
\nsc{
Our derivation relies on three central assumptions: (i) the Oberbeck-Boussinesq approximation can be applied to the liquid phase, while the gas thermophysical properties are a generic function of thermodynamic pressure, local temperature and vapour composition, (ii) the Grossmann-Lohse theory for thermal convection can be applied to the liquid and gas layers separately, (iii) the vapour content in the gas can be taken as the mean value at the gas-liquid interface. The validity of these assumptions is assessed and discussed in \S~\ref{sec:assess}.
}
%
%
\subsection{Interface temperature}\label{sec:theta_g}
\nsc{
The first step is to account for the variation of the liquid height induced by phase change and for the NOB effects \cite[relevant for $\varepsilon\geq 0.05$, see][]{chilla2012new,wan2020non} due to variations of the local temperature, thermodynamic pressure and composition. Note that based on the first assumption of our derivation, NOB effects manifest only in the gas phase, whereas the liquid phase is described under the OB approximation. For simplicity of notation, a generic quantity $\widehat{\xi}_g$ is expressed as
\begin{equation}
  \widehat{\xi}_g = \widehat{\xi}_{g,r}\left(1+\dfrac{\Delta\widehat{\xi}_g}{\widehat{\xi}_{g,r}}\right)=\widehat{\xi}_{g,r}f_{g,\xi}\mathrm{,}\label{eqn:xi_g}
\end{equation}
where $\Delta\widehat{\xi}_g=\widehat{\xi}_g-\widehat{\xi}_{g,r}$ and $f_{g,\xi}$ is the mean normalized variation of $\widehat{\xi}_g$ with respect to the same quantity evaluated at reference condition, $\widehat{\xi}_{g,r}$. Both $\widehat{\xi}_{g,r}$ and $f_{g,\xi}$ refer to the layer pertaining to the gas phase. Following the approach proposed by~\cite{liu2021heat} and using~\eqref{eqn:xi_g}, we define a Rayleigh number in each phase,
\begin{equation}
  \begin{cases}
  Ra_l = \dfrac{\widehat{\beta}_l(\widehat{T}_b-\widehat{T}_{\Gamma})|\widehat{\mathbf{g}}|\widehat{h}_l^3\widehat{\rho}_l^2\widehat{c}_{pl}}{\widehat{\mu}_l\widehat{k}_l}=\alpha_0^3(1/2-\Theta_{\Gamma})Ra\dfrac{\lambda_{\rho}^2\lambda_{cp}\lambda_{\beta}}{\lambda_{\mu}\lambda_k}f_{l,h}^3\mathrm{,} \\
  Ra_g = \dfrac{\widehat{\beta}_g(\widehat{T}_{\Gamma}-\widehat{T}_t)|\widehat{\mathbf{g}}|\widehat{h}_g^3\widehat{\rho}_g^2\widehat{c}_{pg}}{\widehat{\mu}_g\widehat{k}_g}=(1-\alpha_0)^3(1/2+\Theta_{\Gamma})Ra\dfrac{f_{g,h}^3f_{g,\rho}^2f_{g,cp}}{f_{g,\mu}f_{g,k}}\mathrm{,}
  \end{cases}
  \label{eqn:ral_per_phase}
\end{equation}
\sloppy
where $\widehat{h}_i$ is the height of each layer, $\widehat{\beta}_{i}$ is the thermal expansion coefficient, $\widehat{\rho}_{i}$, $\widehat{\mu}_{i}$, $\widehat{k}_{i}$ and $\widehat{c}_{p,i}$ are the fluid density, dynamic viscosity, conductivity and specific heat capacity and $\lambda_{\xi}$ the associated property ratio scaled with respect to the reference gas property evaluated at $\widehat{T}_r$, $\widehat{p}_{th,r}$ and in a dry condition, i.e.\ $Y_l^v=0$. Further, $|\widehat{\mathbf{g}}|$ is the gravity acceleration, $\alpha_0$ is the initial liquid volume fraction and $Ra=\widehat{\beta}_g\widehat{\Delta T}|\widehat{\mathbf{g}}|\widehat{l}_z^3\widehat{\rho}_{g,r}^2\widehat{c}_{pg,r}/(\widehat{\mu}_{g,r}\widehat{k}_{g,r})$ is a "fictitious" Rayleigh number based on the reference gas thermophysical properties, the height of the cavity and the temperature difference between top and bottom walls. Given the NOB effects in the gas, we define $\widehat{\beta}_g=1/\widehat{T}_{c,g}$, while $\widehat{\beta}_l$ is taken as constant, which is consistent with the OB approximation. $\widehat{T}_{c,g}$ is the central temperature in the gas region and its estimation is given later in this section. \par
Next, we define two separate Nusselt numbers, $Nu_l$ and $Nu_g$,
\begin{equation}
  \begin{cases}
  Nu_l = \dfrac{\widehat{Q}_{\Gamma,l}\widehat{h}_l}{\widehat{k}_l(\widehat{T}_b-\widehat{T}_\Gamma)}= \dfrac{\widehat{Q}_{\Gamma,l}\alpha_0\widehat{l}_zf_{l,h}}{\widehat{k}_{l}(1/2-\Theta_{\Gamma})\widehat{\Delta T}}\mathrm{,} \\
  Nu_g = \dfrac{\widehat{Q}_{\Gamma,g}\widehat{h}_g}{\widehat{k}_g(\widehat{T}_\Gamma-\widehat{T}_t)} = \dfrac{\widehat{Q}_{\Gamma,g}(1-\alpha_0)\widehat{l}_zf_{g,h}}{\widehat{k}_{g,r}f_{g,k}(1/2+\Theta_{\Gamma})\widehat{\Delta T}}\mathrm{,}
  \end{cases}
  \label{eqn:nus_per_phase}
\end{equation}
where $\widehat{Q}_{\Gamma,l}$ and $\widehat{Q}_{\Gamma,g}$ are the heat fluxes on the liquid and gas side of the interface. In the absence of phase change and when evaporation and condensation events are statistically balanced, these two quantities are on average equal. Employing the GL theory, the Nusselt number in both layers is then related to the corresponding Rayleigh number. Note that the complete GL theory is a system of equations that provides the value of the Nusselt $Nu$ and the Reynolds $Re$ numbers for a given Rayleigh $Ra$ and Prandtl $Pr$ numbers. The implicit nature of this system prevents obtaining an explicit expression for $\Theta_\Gamma$ and, therefore, the following simplified scaling laws are here considered~\citep{weiss2018bulk},
\begin{equation}
  Nu_l = A_lRa_l^{\gamma_l} Pr_l^{m_l}, \hspace{1 cm} Nu_g = A_gRa_g^{\gamma_g} Pr_g^{m_g}
  \label{eqn:simp_gl}
\end{equation}
In this work, we consider $A=A_l=A_g$, $\gamma=\gamma_l=\gamma_g$ and $m=m_l=m_g$. As remarked in~\cite{weiss2018bulk}, employing the simplified GL theory in equations~\eqref{eqn:simp_gl} is valid as long as $Ra_l$, $Ra_g$ and $Pr_l$, $Pr_g$ are sufficiently similar to fall inside the same scaling regime~\citep{grossmann2000scaling,grossmann2001thermal} so that the same $\gamma$ and $m$ can be used for both layers. Taking the ratio $Nu_l/Nu_g$ yields
\begin{equation}
  \dfrac{Nu_l}{Nu_g} = \left(\dfrac{Ra_l}{Ra_g}\right)^\gamma\left(\dfrac{Pr_l}{Pr_g}\right)^m\mathrm{.}
  \label{eqn:simp_gl_ratio}
\end{equation}
As suggested in~\cite{weiss2018bulk}, if $Pr>0.5$ the GL theory suggests a scaling exponent $m$ very close to zero and the Prandtl dependence in equation~\eqref{eqn:simp_gl_ratio} can be omitted. To obtain an explicit relation for $\Theta_\Gamma$, we compute the ratio $Ra_l/Ra_g$ from equations~\eqref{eqn:ral_per_phase}
\begin{equation}
  \dfrac{Ra_l}{Ra_g} = \dfrac{\alpha_0^3}{(1-\alpha_0)^3}\dfrac{f_{l,h}^3}{f_{g,h}^3}\dfrac{(1/2-\Theta_\Gamma)}{(1/2+\Theta_\Gamma)}\dfrac{\mathcal{F}_\lambda}{f_{g,\rho}^2\mathcal{F}_g}\dfrac{\lambda_\beta}{\lambda_k}f_{g,k}\mathrm{,}
  \label{eqn:ratio_ra}
\end{equation}
where $\mathcal{F}_\lambda=\lambda_{\rho}^2\lambda_{cp}/\lambda_\mu$ and $\mathcal{F}_g=f_{g,cp}/f_{g,\mu}$. Likewise, we express $Nu_l/Nu_g$ using equations~\eqref{eqn:nus_per_phase} as
\begin{equation}
  \dfrac{Nu_l}{Nu_g} = \dfrac{\alpha_0}{(1-\alpha_0)}\dfrac{f_{l,h}}{f_{g,h}}\dfrac{1}{\lambda_k}\dfrac{(1/2+\Theta_\Gamma)}{(1/2-\Theta_\Gamma)}f_{g,k}\mathrm{.}
  \label{eqn:ratio_nu}
\end{equation}
Note that to derive equation~\eqref{eqn:ratio_nu}, the heat flux on the liquid and gas side are taken equal $\widehat{Q}_{\Gamma,l}=\widehat{Q}_{\Gamma,g}$. Once more, this is a valid assumption when evaporation and condensation balances at the interface and the gas layer is at saturation. By employing the scaling relations~\eqref{eqn:simp_gl} and equations~\eqref{eqn:ratio_ra} and~\eqref{eqn:ratio_nu}, we get:
\begin{align}
  &\dfrac{\alpha_0}{(1-\alpha_0)}\dfrac{f_{l,h}}{f_{g,h}}\dfrac{1}{\lambda_k}f_{g,k}\dfrac{(1/2+\Theta_\Gamma)}{(1/2-\Theta_\Gamma)} = \\
  &\left[\dfrac{\alpha_0^3}{(1-\alpha_0)^3}\dfrac{f_{l,h}^3}{f_{g,h}^3}\dfrac{(1/2-\Theta_\Gamma)}{(1/2+\Theta_\Gamma)}\dfrac{\mathcal{F}_\lambda}{f_{g,\rho}^2\mathcal{F}_g}\dfrac{\lambda_\beta}{\lambda_k}f_{g,k}\right]^\gamma\mathrm{,}\nonumber 
\end{align}
which, after some manipulation, reads
\begin{equation}
  \dfrac{1}{\Theta_\Gamma^*} = 1+\left(\dfrac{\alpha_0}{1-\alpha_0}\dfrac{f_{l,h}}{f_{g,h}}\right)^{\dfrac{1-3\gamma}{1+\gamma}}\left(\dfrac{f_{g,\rho}^2\mathcal{F}_g}{\mathcal{F}_\lambda\lambda_\beta}\right)^{\dfrac{\gamma}{1+\gamma}}\left(\dfrac{f_{g,k}}{\lambda_k}\right)^{\dfrac{1-\gamma}{1+\gamma}}\mathrm{.}
  \label{eqn:theta_g_temp}
\end{equation}
Note that in the derivation of equation~\eqref{eqn:theta_g_temp}, we have performed a change of variable, $\Theta_\Gamma^*=\Theta_\Gamma+1/2$. Equation \eqref{eqn:theta_g_temp} can be then rearranged in terms of $\Theta_\Gamma^*$ and, finally, of $\Theta_\Gamma$,
\begin{equation}
  \Theta_{\Gamma} = -\dfrac{1}{2}+\left(1+\left(\dfrac{\alpha_0}{1-\alpha_0}\dfrac{f_{l,h}}{f_{g,h}}\right)^{\dfrac{1-3\gamma}{1+\gamma}}\left(\dfrac{f_{g,\rho}^2\mathcal{F}_g}{\mathcal{F}_{\lambda}\lambda_{\beta}}\right)^{\dfrac{\gamma}{1+\gamma}}\left(\dfrac{f_{g,k}}{\lambda_k}\right)^{\dfrac{1-\gamma}{1+\gamma}}\right)^{-1}\mathrm{,}
  \label{eqn:theta_g}
\end{equation}
where $\lambda_\beta$ is defined as $\lambda_\beta=\widehat{\beta}_l/\widehat{\beta}_g$ with $\widehat{\beta}_g=1/\widehat{T}_{c,g}$. Accordingly, $\lambda_\beta$ can be finally expressed as
\begin{equation}
  \lambda_\beta = \widehat{\beta}_l\widehat{T}_{c,g}=\widehat{\beta}_l\widehat{\Delta T}\left(\dfrac{\widehat{T}_{c,g}-\widehat{T}_r}{\widehat{\Delta T}}+\dfrac{\widehat{T}_r}{\widehat{\Delta T}}\right)=\widehat{\beta}_l\widehat{T}_r\left(1+2\varepsilon\Theta_c\right)\mathrm{.}
\end{equation}
In equation~\eqref{eqn:theta_g}, $f_{g,k}$ and $\mathcal{F}_{g}$ are function of temperature and composition. The dependence on the thermodynamic pressure is typically important for the density, i.e.\ $f_{g,\rho}$, while it can be omitted for the specific heat capacity, viscosity, and thermal conductivity. Therefore, to evaluate $f_{g,k}$ and $\mathcal{F}_{g}$ in~\eqref{eqn:theta_g}, we need to specify a reference temperature and reference vapor concentration. This aspect has already been discussed in~\cite{weiss2018bulk}, where the authors derive an expression for the central temperature for the gas region $\widehat{T}_{c,g}$ and show that it represents the temperature at which the thermophysical properties should be evaluated for a Rayleigh-B\'enard cell under strong NOB effects. Note that the derivation is based on the same assumptions that lead to equations~\eqref{eqn:simp_gl} and~\eqref{eqn:simp_gl_ratio} and, therefore, no additional hypotheses are introduced here. By incorporating this approach in our model, $\widehat{T}_{c,g}$ reads as
\begin{equation}
  \widehat{T}_{c,g} = \dfrac{\widehat{k}_{+,g}^{3/4}\widehat{\eta}_{+,g}^{1/4}\widehat{T}_\Gamma+\widehat{k}_{-,g}^{3/4}\widehat{\eta}_{-,g}^{1/4}\widehat{T}_t}{\widehat{k}_{+,g}^{3/4}\widehat{\eta}_{+,g}^{1/4}+\widehat{k}_{-,g}^{3/4}\widehat{\eta}_{-,g}^{1/4}}\mathrm{.}
  \label{eqn:t_c_g}
\end{equation}
Note that when the gas thermophysical properties are uniform, $\widehat{T}_{c,g}$ reduces to $(\widehat{T}_\Gamma+\widehat{T}_t)/2$ and, therefore, equation~\eqref{eqn:t_c_g} can be interpreted as a more general choice of the central temperature than the arithmetic mean. By introducing $\Theta_{c,g}=(\widehat{T}_{c,g}-\widehat{T}_r)/\widehat{\Delta T}$, equation~\eqref{eqn:t_c_g} can be written in dimensionless form as
\begin{equation}
  \Theta_{c,g} = \dfrac{\Theta_\Gamma-\dfrac{1}{2}\dfrac{\widehat{k}_{-,g}^3\widehat{\eta}_{-,g}}{k_{+,g}^3\widehat{\eta}_{+,g}}}{1+\dfrac{1}{2}\dfrac{\widehat{k}_{-,g}^3\widehat{\eta}_{-,g}}{\widehat{k}_{+,g}^3\widehat{\eta}_{+,g}}}\mathrm{.}
  \label{eqn:theta_c}
\end{equation}
Note that in equations~\eqref{eqn:t_c_g} and~\eqref{eqn:theta_c}, the group $\widehat{\eta}_{\pm,g}$ corresponds to $(\widehat{\beta}\widehat{\rho}^2/\widehat{k}\widehat{\mu})_{\pm,g}$. The properties $\widehat{k}_{+,g}$, $\widehat{\eta}_{+,g}$ and $\widehat{k}_{-,g}$, $\widehat{\eta}_{-,g}$ are evaluated at the crossover points, which are located at the transition points between the boundary layer and the bulk region of the cell. Following once more the procedure in~\cite{weiss2018bulk}, the crossover temperature $\widehat{T}_{\pm,g}$ at which these properties should be evaluated is determined as a linear combination between $\widehat{T}_{\Gamma}$ and $\widehat{T}_t$. In particular, for the gas region we have:
\begin{equation}
  \widehat{T}_{+,g} = \delta \widehat{T}_\Gamma + (1-\delta)\widehat{T}_{c,g}\mathrm{,} \hspace{0.5 cm} \widehat{T}_{-,g} = \delta \widehat{T}_t + (1-\delta)\widehat{T}_{c,g}\mathrm{.}
\end{equation}
The last parameter to be specified is $\delta$, which depends exponentially on the aspect ratio $\mathcal{A}$, i.e.\ $\delta = \delta_0 + (1-\delta_0)\exp{(-B\mathcal{A})}$. By experimental fitting, \cite{weiss2018bulk} suggest to employ $\delta_0=0.235$ and $B=1.14$, which provide an accurate estimation of $\Theta_{c,i}$ for a wide range of $Ra$ and $\mathcal{A}$.
\par
For the terms $f_{g,\rho}$, $f_{l,h}$ and $f_{g,h}$, more analysis is needed. Since evaporation changes the mass of the gas and its local density, and it decreases the volume of the liquid, we introduce the mass ratio $G_g=\widehat{G}_g/\widehat{G}_{g,r}$ and the volume ratio $V_g=\widehat{V}_g/\widehat{V}_{g,r}$. These ratios are defined as the values of the quantities in the evaporating regime  
divided by the corresponding values for the flow without evaporation. Indicating with $\widehat{G}_{l,r}$ the initial liquid mass, $\widehat{V}_{l,r}$ the initial liquid volume and $\widehat{G}_l^e$ the mass of the evaporated liquid, we get
\begin{equation}
  \begin{cases}
  \dfrac{\widehat{G}_g}{\widehat{G}_{g,r}} = \dfrac{\widehat{G}_{g,r}+\widehat{G}_{l,r}-\widehat{G}_l}{\widehat{G}_{g,r}} = 1 + \dfrac{\widehat{G}^e_{l}}{\widehat{G}_{g,r}} = 1 + \dfrac{1}{\widehat{G}_{g,r}}\displaystyle{\int_{\widehat{V}_g}\widehat{\rho}_gY_l^v\widehat{dV}_g}\mathrm{,} \\
  \dfrac{\widehat{V}_g}{\widehat{V}_{g,r}} = \dfrac{\widehat{V}_{g,r}+\widehat{V}_{l,r}-\widehat{V}_l}{\widehat{V}_{g,r}} = 1 + \dfrac{\widehat{V}_l^e}{\widehat{V}_{g,r}} = 1 + \dfrac{1}{\lambda_{\rho}\widehat{G}_{g,r}}\displaystyle{\int_{\widehat{V}_g}\widehat{\rho}_gY_l^v\widehat{dV}_g}\mathrm{,}
  \end{cases}
  \label{eqn:mas_vol_g}
\end{equation} 
where $\widehat{V}_l^e$ is the volume of the incompressible liquid that turns into vapour. Note that $\widehat{G}_l^e$, given by the integrals in equations~\eqref{eqn:mas_vol_g}, requires knowledge of the vapour distribution. Approximating $Y_l^v$ with $\overline{Y}_{l,\Gamma}^v$ (second assumption of our derivation) allows us to estimate the mass of the evaporated liquid as $\widehat{G}_l^e\approx\overline{Y}_{l,\Gamma}^v\widehat{G}_g$. Accordingly, $G_g=1/(1-\overline{Y}_{l,\Gamma}^v)$ and $V_g=1+\overline{Y}_{l,\Gamma}^v/(\lambda_{\rho}(1-\overline{Y}_{l,\Gamma}^v))$, which provide the following estimation of $f_{g,\rho}$:
\begin{equation}
  f_{g,\rho} = \dfrac{G_g}{V_g} = \dfrac{\lambda_{\rho}}{\overline{Y}_{l,\Gamma}^v+(1-\overline{Y}_{l,\Gamma}^v)\lambda_{\rho}}\mathrm{.}
  \label{eqn:rhge}
\end{equation}
Since $f_{i,h}\approx \widehat{G}_i/(\widehat{\rho}_i\widehat{V}_{i,r})$, the relations~\eqref{eqn:mas_vol_g} allow also to estimate $f_{l,h}$ and $f_{g,h}$. \par
To proceed, it is worth noticing that $\overline{Y}_{l,\Gamma}^v=g_1(\Pi_T,\Pi_P,\varepsilon,\Theta_{\Gamma},p_{th})$ from eq.~\eqref{eqn:rault_sw_par} and $\Theta_{\Gamma}=g_2(\Pi_T,\Pi_P,\varepsilon,p_{th})$ as in~\eqref{eqn:theta_g1_yes_eva} and, therefore, a relation for the thermodynamic pressure \cite[supposed uniform, see][]{chilla2012new} is required. To derive it, we integrate over the gas region the equation of state for the local gas density, i.e.\ $p_{th}M_{m}/(1+2\varepsilon\Theta_g)$ and express the result in terms of $p_{th}$,
\begin{equation}
  p_{th} = {G_g}\left({\displaystyle{\int_{V_g}M_{m}\Theta_g^idV_g}}\right)^{-1}\hspace{-0.25 cm}
  \approx \dfrac{f_{g,\rho}}{\overline{M}_{m}}\Theta_c^i \, \mathbf{,}
  \label{eqn:pth_ratio}
\end{equation}
where $\Theta_g^i=(1+2\varepsilon\Theta_g)^{-1}$ and $\overline{M}_{m}=\lambda_M/(\overline{Y}_{l,\Gamma}^v+(1-\overline{Y}_{l,\Gamma}^v)\lambda_M)$ is the mean molar mass of the mixture computed using the harmonic average between $\widehat{M}_l$ and $\widehat{M}_{g,r}$~\citep{scapin2022finite}. Note that in equation~\eqref{eqn:pth_ratio}, we employ the second hypothesis and we approximate the volume integral of $\Theta_g^i$ as $V_g\overline{\Theta}_g^i$ using $\overline{\Theta}_g^i=\Theta_c^i$ from equation~\eqref{eqn:theta_c}. \par
The model to estimate $\Theta_\Gamma$ is based on equations~\eqref{eqn:theta_g},~\eqref{eqn:theta_c},~\eqref{eqn:rhge} and~\eqref{eqn:pth_ratio}, coupled with appropriate equations of state for specific heat capacity, thermal conductivity, and viscosity, as detailed in Appendix~\ref{sec:eos}. The system is not linear; however, a simple iterative procedure can be used to obtain $\Theta_\Gamma$, $\overline{Y}_l^v$ and $p_{th}$ together with $\Theta_{c,g}$. We want to remark here that the proposed model for $\Theta_\Gamma$ is more general than the one presented in~\cite{liu2021heat} on three aspects: i) we account for phase change, ii) we account for NOB effects in the gas phase, iii) we include the density, viscosity, specific heat capacity and thermal expansion ratios in equation~\eqref{eqn:theta_g}. It is worth mentioning that in the absence of evaporation, with uniform bulk properties (i.e.\ $\mathcal{F}_g=f_{g,\xi}=1$) and for $\lambda_\rho=\lambda_\mu=\lambda_{cp}=\lambda_{\beta}=1$, the general expression for $\Theta_\Gamma$ in equation~\eqref{eqn:theta_g} reduces to the estimate by~\cite{liu2021heat}.
}
%
%
\subsection{Nusselt number}\label{sec:nuss_eva}
\nsc{
With the estimated interface temperature $\Theta_{\Gamma}$, we can derive a scaling law for the ratio between the global Nusselt number with and without evaporation, i.e.\ $Nu^e=\widehat{Q}_t^e\widehat{l}_z/(\widehat{k}_{g,r}\widehat{\Delta T})$ and $Nu=\widehat{Q}_t\widehat{l}_z/(\widehat{k}_{g,r}\widehat{\Delta T})$, where $\widehat{Q}_t^e$ and $\widehat{Q}_t$ are the global heat fluxes with and without evaporation measured at the top boundary. Taking the ratio between the two, we immediately see that $Nu^e/Nu = \widehat{Q}_t^e/\widehat{Q}_t$. To compute $\widehat{Q}_t^e/\widehat{Q}_t$, we first define the Nusselt number on the gas side of the interface, with and without evaporation,
\begin{equation}
  Nu_g^e = \dfrac{\widehat{Q}_{\Gamma}^e\widehat{h}_g^e}{\widehat{k}_g^e(\Theta_{\Gamma}^e+1/2)\widehat{\Delta T}}\mathrm{,} \quad   Nu_{g} = \dfrac{\widehat{Q}_{\Gamma}\widehat{h}_g}{\widehat{k}_{g}(\Theta_{\Gamma}+1/2)\widehat{\Delta T}}\mathrm{,}
  \label{eqn:nu_ge}
\end{equation}
where $\widehat{Q}_{\Gamma}^e$ and $\widehat{Q}_{\Gamma}$ are the heat fluxes exchanged at the interface. Taking the ratio of the two Nusselt numbers in equation~\eqref{eqn:nu_ge} and applying again the GL theory (i.e.\ $Nu_g^e/Nu_g=\left(Ra_g^e/Ra_g\right)^{\gamma}$) yields 
\begin{equation}
  \dfrac{\widehat{Q}_{\Gamma}^e}{\widehat{Q}_{\Gamma}} = \dfrac{(\Theta_{\Gamma}^e+1/2)}{(\Theta_{\Gamma}+1/2)}\left(\dfrac{Ra_{g}^e}{Ra_g}\right)^{\gamma}\dfrac{f_{g,k}^e}{f_{g,k}}\dfrac{1}{f_{g,h}^e}\mathrm{.}
  \label{eqn:qte_qt}
\end{equation}
Note that the variations of the thermophysical properties induced by evaporation are denoted with a superscript "e". We then use equation~\eqref{eqn:ral_per_phase} to compute the ratio
\begin{equation}
  \dfrac{Ra_g^e}{Ra_{g}} = \dfrac{(1+2\varepsilon\Theta_c^e)}{(1+2\varepsilon\Theta_c)}\dfrac{(\Theta_{\Gamma}^e+1/2)}{(\Theta_{\Gamma}+1/2)}f_{g,h}^{3,e}f_{g,\rho}^{2,e}\dfrac{f_{g,cp}^{e}}{f_{g,\mu}^{e}f_{g,k}^{e}}\dfrac{f_{g,\mu}f_{g,k}}{f_{g,cp}}\mathrm{.}
  \label{eqn:ratio_rag}
\end{equation}
Since the global heat flux is equal to the heat flux at the interface, i.e.\ $\widehat{Q}_{\Gamma}^e=\widehat{Q}_{t}^e$ and $\widehat{Q}_{\Gamma}=\widehat{Q}_{t}$, we can finally combine equations~\eqref{eqn:qte_qt} and~\eqref{eqn:ratio_rag} into
\begin{equation}
  \dfrac{Nu^e}{Nu} = \left(\dfrac{1+2\varepsilon\Theta_c}{1+2\varepsilon\Theta_c^e}\right)^{\gamma}\left(\dfrac{\Theta_{\Gamma}^e+1/2}{\Theta_{\Gamma}+1/2}\right)^{1+\gamma}\dfrac{f_{g,\rho}^{2\gamma,e}}{f_{g,h}^{1-3\gamma,e}}\dfrac{f_{g,cp}^{\gamma,e}}{f_{g,cp}^{\gamma}}\dfrac{f_{g,\mu}^{\gamma}}{f_{g,\mu}^{e,\gamma}}\dfrac{f_{g,k}^{1-\gamma,e}}{f_{g,k}^{1-\gamma}}\mathrm{.}
  \label{eqn:qte_qt_1}
\end{equation}
Solving iteratively the system composed of equations~\eqref{eqn:theta_g} and~\eqref{eqn:pth_ratio}, together with the relations~\eqref{eqn:rhge} and~\eqref{eqn:rault_sw_par}, provides the values of $\Theta_\Gamma$ and $p_{th}$. Once these are known, we can directly estimate the global heat transfer modulation with equation~\eqref{eqn:qte_qt_1}. This expression predicts $Nu$ for an evaporating system with respect to the configuration without phase change, described by the GL theory. Inspection of equation~\eqref{eqn:qte_qt_1} allows us to draw some initial conclusions on the role of phase change on the Rayleigh-B\'enard system. First, the change in liquid height influences the heat transfer modulation only weakly, since its exponent is $1-3\gamma \approx 0$. Second, higher gas density decreases the interface temperature, as suggested by equation~\eqref{eqn:theta_g1_yes_eva}. Nevertheless, despite in equation~\eqref{eqn:qte_qt_1} the exponent of $\Theta_\Gamma$ is larger than the exponent of $f_{g,\rho}$, this last term is expected to be dominant given its stronger dependence on $\varepsilon$, as it is clearly shown in the next section. Last, a non-negligible effect is present due to the variation of $c_p$, $\mu$ and $k$ whose contribution to the heat transfer modulation scales with $\gamma$ and $1-\gamma$. 
}
%
%
\section{Numerical methodology}\label{sec:math_num}
\subsection{Governing equations}
The validation of the model previously described is performed with the in-house code for phase-changing flows extensively described in~\cite{scapin2020volume,scapin2022finite} and, therefore, we briefly mention here only the main features. First, to distinguish between the phases, an indicator function $H$ is introduced and it is defined equal to $1$ in the liquid and $0$ in the gas phase. $H$ is governed by the following transport equation
\begin{equation}
  \dfrac{\partial H}{\partial t} + \mathbf{u}_\Gamma\cdot\nabla H = 0\mathrm{,}
  \label{eqn:h_ind}
\end{equation}
where $\mathbf{u}_\Gamma$ is the interface velocity computed as the sum of an extended liquid velocity, $\mathbf{u}_l^e$, and a term due to phase change, $\dot{m}_\Gamma/\rho_l\mathbf{n}_\Gamma$, with $\dot{m}_\Gamma$ the mass flux and $\mathbf{n}_\Gamma$ the unit normal vector at the interface. Note that $\mathbf{u}_l^e$ is computed as described in~\cite{scapin2020volume}. Next, the numerical code solves the governing equations assuming that the liquid phase has constant properties and can be treated within the Oberbeck-Boussinesq (OB) approximation, while the gas phase manifests compressible effects that can be described within the low-Mach number formulation. Accordingly, the dimensionless conservation equations for momentum, vaporized species $Y_l^v$, temperature $\Theta$ and mass-flux $\dot{m}_{\Gamma}$ across the interface read~\citep{scapin2022finite},
\begin{equation}
  \rho\dfrac{D\mathbf{u}}{Dt} =-\nabla p+\sqrt{\dfrac{Pr}{Ra}}\nabla\cdot\mathbf{\tau}+\dfrac{\mathbf{f}_{\sigma}}{We}+\left[\lambda_{\rho}\left(1-\Pi_{\beta}2\varepsilon\Theta\right)H+\dfrac{p_{th}M_{m}}{(1+2\varepsilon\Theta)}(1-H)\right]\dfrac{\mathbf{e}_z}{2\varepsilon}\mathrm{,} \label{eqn:mom1}
\end{equation}
\begin{equation}
  \rho_g\dfrac{DY_{l}^{v}}{Dt} = \dfrac{\nabla\cdot(\rho_gD_{lg}\nabla Y_{l}^{v})}{\sqrt{RaSc}} \label{eqn:vap1}\mathrm{,}
\end{equation}
\begin{equation}
  \rho c_p\dfrac{D\Theta}{Dt} = \dfrac{\nabla\cdot(k\nabla\Theta)}{\sqrt{RaPr}} + \Pi_{R}\dfrac{dp_{th}}{dt}(1-H) - \dfrac{(\dot{m}_{\Gamma}\delta_{\Gamma})}{2\varepsilon Ste}\mathrm{,} \label{eqn:tmp1}
\end{equation}
\begin{equation}
  \dot{m}_{\Gamma} = \dfrac{1}{\sqrt{RaSc}}\dfrac{\rho_{g,\Gamma}D_{lg,\Gamma}}{1-Y_{l,\Gamma}^v}\nabla_{\Gamma} Y_{l}^{v}\cdot\mathbf{n}_{\Gamma}\label{eqn:int_mfx}\mathrm{.}
\end{equation}
In~\eqref{eqn:mom1}, $\mathbf{u}$ is the velocity, $p$ is the hydrodynamic pressure, $\tau$ is the viscous stress tensor for compressible Newtonian flows and $\mathbf{f}_{\sigma}=\kappa_{\Gamma}\delta_{\Gamma}\mathbf{n}_\Gamma$ with $\kappa_{\Gamma}$ the interfacial curvature, $\delta_\Gamma$ a regularized Dirac-delta function~\citep{scardovelli1999direct} and $\mathbf{n}_\Gamma$ the normal vector. The unit vector $\mathbf{e}_z$ points in the gravity direction, i.e.\ $\mathbf{e}_z=(0,0,-1)$. \par
The generic thermophysical property $\xi$ (density $\rho$, dynamic viscosity $\mu$, thermal conductivity $k$ or specific heat capacity $c_p$) is computed with an arithmetic average, i.e.\ $\xi=1+(\lambda_{\xi}-1)H$ where $\lambda_{\xi}=\xi_l/\xi_{g,r}$ with $\xi_{g,r}$ evaluated at reference condition (i.e.\ $\widehat{T}_r$, $\widehat{p}_{th,r}$ and $Y_l^v=0$). Since $\xi_l$ is kept constant and uniform no further modeling is needed, while the generic gas property $\xi_{g}$ is computed with the appropriate equation of state. For example, the gas density is computed with the ideal gas law, $\rho_g=p_{th}M_{m}/(1+2\varepsilon\Theta)$, and the vapour diffusion coefficient with the Wilke-Lee correlation~\citep{reid1987properties}, $D_{lg}=(1+2\varepsilon\Theta)^{3/2}/p_{th}$~\citep{wan2020non}. The remaining gas thermophysical properties are computed as detailed in the appendix~\ref{sec:eos}. Note that in the Oberbeck-Boussinesq limit, (i.e.\ $\varepsilon\rightarrow 0$, $p_{th}=1$ and $M_{m}=1$), the gravity term active in the gas region reduces to $1-2\varepsilon\Theta$ with $\widehat{\beta}_g=1/\widehat{T}_r$ and hence matches the one employed in previous works~\citep{liu2021efficient,liu2021two,liu2022enhancing,liu2021heat}. \par
\nsc{
Equations~\eqref{eqn:mom1},~\eqref{eqn:vap1},~\eqref{eqn:tmp1} and \eqref{eqn:int_mfx} are written in dimensionless form by introducing the free-fall velocity scale $\widehat{u}_r=(2\varepsilon|\widehat{\mathbf{g}}|\widehat{l}_z)^{1/2}$ and the free-fall timescale $\widehat{t}_r=\widehat{l}_z/\widehat{u}_r$. Accordingly, we define the Weber number $We=\widehat{\rho}_{g,r}2\varepsilon|\widehat{\mathbf{g}}|\widehat{l}_{z}^2/\widehat{\sigma}$ with $\widehat{\sigma}$ the surface tension, $\Pi_{\beta}=\widehat{\beta_l}\widehat{T_r}$; $Sc=\widehat{\mu}_{g,r}/(\widehat{\rho}_{g,r}\widehat{D}_{lg,r})$ and $Pr=\widehat{\mu}_{g,r}\widehat{c}_{pg,r}/\widehat{k}_{g,r}$ are the Schmidt and the Prandtl numbers. Based on the chosen reference quantities, the Rayleigh number $Ra=2\varepsilon|\widehat{\mathbf{g}}|\widehat{l}_z^3\widehat{\rho}_{g,r}^2\widehat{c}_{pg,r}/(\widehat{\mu}_{g,r}\widehat{k}_{g,r})$ corresponds to the fictitious one defined in equations~\eqref{eqn:ral_per_phase}. Note that the temperature equation~\eqref{eqn:tmp1} requires the definition of the Stefan number $Ste=\widehat{c}_{pg,r}\widehat{T}_r/\widehat{\Delta h}_{lv}$, where $\widehat{\Delta h}_{lv}$ is the latent heat, and of the dimensionless group $\Pi_{R}=\widehat{R}_u/(\widehat{c}_{pg,r}\widehat{M}_{g,r})$, where $\widehat{M}_{g,r}$ is the molar mass of the gas phase and $\widehat{R}_u$ the universal gas constant. In equation~\eqref{eqn:int_mfx}, the vapor mass fraction at the interface $Y_{l,\Gamma}^v$ is computed using equations~\eqref{eqn:rault_sw_par}. These are the Raoult's law~\citep{reid1987properties} and the Span-Wagner equation of state for the vapour pressure $\widehat{p}_{s,\Gamma}$ at the gas-liquid interface. The employed coefficients are those for pentane and taken equal to $B_{i=1,4}=[-7.327140,+1.823650,-2.272744,-2.711929]$. 
}
Note that we prefer to employ the slightly more elaborated Span-Wagner model over "simpler" equations for the partial pressure, e.g. the Clausius-Clapeyron's or Antoine's laws~\citep{reid1987properties}. The motivation behind our choice is twofold. First, the Span-Wagner equation is based on critical quantities, $\widehat{T}_{cr}$ and $\widehat{p}_{cr}$, which are intrinsic properties of the substance and thus independent of the local ambient conditions (e.g., $\widehat{p}_{th,r}$). Next, the Span-Wagner model provides accurate and reliable results for most of the substances over a wide range of temperature and thermodynamic pressure, well below the critical point as well as near it. \par
To form a close set of equations, one needs a relation for the velocity divergence and for the thermodynamic pressure,
\begin{equation}
  \nabla\cdot\mathbf{u} = f_{\Gamma} + \left[\dfrac{p_{th}f_{Y} + f_\Theta}{p_{th}} - \left(1-\dfrac{\Pi_{R}}{c_pM_{m}}\right)\dfrac{1}{p_{th}}\dfrac{dp_{th}}{dt}\right](1-H)
  \label{eqn:veldiv1}\mathrm{,}
\end{equation}
\begin{equation}
  \dfrac{1}{p_{th}}\dfrac{dp_{th}}{dt}\int_{V_g}\left(1-\dfrac{\Pi_{R}}{c_pM_{m}}\right)dV_g = \int_V\left[f_{\Gamma}+\dfrac{f_\Theta+p_{th}f_Y}{p_{th}}(1-H)\right] dV
  \label{eqn:pth1}\mathrm{.}
\end{equation}
\nsc{
The local divergence constrain, equation~\eqref{eqn:veldiv1}, is derived by applying the divergence operator to the one-fluid velocity defined as $\mathbf{u}=H\mathbf{u}_l+(1-H)\mathbf{u}_g$ and by applying the continuity equation of both phases. Equation~\eqref{eqn:pth1} is derived by integrating equation~\eqref{eqn:veldiv1} over the total domain $V$, sum of the liquid and gas domains, and by imposing the volume conservation over $V$, i.e.\ $\int_V\nabla\cdot\mathbf{u}dV=0$.} In equations~\eqref{eqn:veldiv1} and~\eqref{eqn:pth1} the functions $f_{\Gamma}$, $f_{Y}$ and $f_{\Theta}$ represent the different contributions to the total velocity divergence from the phase change ($f_{\Gamma}$) and the change of the gas density due to composition ($f_Y$) and temperature ($f_{\Theta}$)~\citep[see again][for the details]{scapin2022finite},
\begin{subequations}
\begin{align}
  f_{\Gamma} &= \dot{m}_{\Gamma}\left(\dfrac{1}{\rho_{g,\Gamma}}-\dfrac{1}{\lambda_{\rho}}\right)\delta_{\Gamma}\mathrm{,} \\
  f_{Y} &= \dfrac{1}{\sqrt{RaSc}}\dfrac{M_{m}}{\rho_g}\left(\dfrac{1}{\lambda_M}-1\right)\nabla\cdot(\rho_gD_{lg}\nabla Y_{l}^{v})\mathrm{,} \\
  f_{\Theta} &= \dfrac{\Pi_{R}}{c_pM_{m}}\dfrac{1}{\sqrt{RaPr}}\nabla\cdot(k\nabla\Theta)\mathrm{.}
\end{align}
  \label{eqn:f_aux}%
\end{subequations}
\nsc{
We remark that a weakly compressible formulation is still required to model an evaporating Rayleigh-B\'enard cell for $\varepsilon\rightarrow 0$. Indeed, the term $f_\Gamma$ is not negligible in case of evaporation and contributes to expansion and contraction at the two-phase interface.} \par 
The governing equations~\eqref{eqn:mom1},~\eqref{eqn:vap1},~\eqref{eqn:tmp1} and~\eqref{eqn:int_mfx} are solved on a uniform Cartesian grid with a standard MAC method~\citep{harlow1965numerical}. The interface dynamics is captured with an algebraic volume-of-fluid method, VoF-MTHINC~\citep{ii2012interface,rosti2019numerical}, which ensures excellent conservation properties of the liquid volume both with and without phase change, provided equation~\eqref{eqn:veldiv1} is correctly imposed on $\mathbf{u}$. Thus, a pressure correction method is employed, and the associated Poisson equation is first factorized into a constant coefficient one~\citep{dodd2014fast} and then solved with the eigenexpansion technique~\citep{schumann1988fast}, as implemented in the open source code CaNS~\citep{costa2018fft}. Further details and validations are provided for phase change problems with constant and variable properties in~\citep{scapin2020volume,dalla2021interface,scapin2022finite}. Appendix~\ref{sec:val} reports an additional validation against the multiphase Rayleigh-B\'enard convection case in \cite{liu2021two}. Note that the baseline two-phase code, including the VoF-MTHINC and heat transfer effects, FluTAS, is described in~\citep{crialesi2023_flutas} and is released as open-source software.
\subsection{Computational setup}
For the validation of the model, we consider three values of the Rayleigh number, $10^6$, $10^7$, and $10^8$, and four values for the temperature differential $\varepsilon=0.05$, $0.10$, $0.15$ and $0.20$. At fixed dimensionless mean temperature $\Pi_T$ and pressure $\Pi_P$, the temperature differential is the only parameter affecting $\Theta_\Gamma$, $Y_l^v$, and $p_{th}$. The remaining dimensionless parameters are not varied; we choose the Prandtl, Schmidt, and Stefan numbers equal to unity, i.e.\ $Pr=Sc=Ste=1$, and the property ratios as $\lambda_{\rho}=\lambda_{\mu}=\lambda_k=20$, $\lambda_{cp}=1$ and $\lambda_M=2.58$. Moreover, we set $\Pi_T=0.8$, $\Pi_{P}=1.65$, $\Pi_{\beta}=0.6$ and $\Pi_{R}=0.18$. This choice corresponds to a light hydrocarbon (e.g., pentane) at high temperature (below the critical value) and high pressure, which can be used as a coolant in industrial applications. \nsc{Finally, we set the Weber number $We=5$ to limit interface deformation for all the investigated values of the Rayleigh number~\citep{liu2021two}.} All the cases are first simulated without evaporation until a statistically stationary condition. Following the procedure proposed for NOB flows~\citep{demou2019direct}, temporal convergence is assessed by comparing the Nusselt number at the bottom and top wall, ensuring that the relative difference is lower than $1$ $\%$. Next, evaporation is activated until a new statistically stationary regime is reached. Also in this case, temporal convergence is assessed by comparing the top and bottom values of $Nu$. First and second-order statistics of the generic variable  $g$ (denoted as $\langle g\rangle_x$ and $\langle g_{rms}\rangle_x$ with $x$ the periodic direction) are collected for a sampling period sufficient to ensure their independence of the size of the sample. Note that we use both Favre and Reynolds averaging. Unless otherwise stated, only the latter is employed in the current work, as we found the difference between the two negligible. In table~\ref{tab:set_up}, we report the sample size and the time step employed for each case with and without evaporation. \par
\begin{table}
\centering
\begin{tabular}{lcccc}
\textit{Case} & $T_{avg}$ - WO EV & $T_{avg}$ - WT EV & $\Delta t_{avg}$ - WO EV & $\Delta t_{avg}$ - WT EV \\
\hline
$Ra=10^6-\varepsilon=0.05$  & $200$  &  $200$ & $5.7\cdot 10^{-5}$ & $2.5\cdot 10^{-5}$  \\
$Ra=10^6-\varepsilon=0.10$  & $200$  &  $200$ & $5.7\cdot 10^{-5}$ & $2.5\cdot 10^{-5}$  \\
$Ra=10^6-\varepsilon=0.15$  & $200$  &  $200$ & $5.7\cdot 10^{-5}$ & $2.5\cdot 10^{-5}$  \\
$Ra=10^6-\varepsilon=0.20$  & $200$  &  $200$ & $5.7\cdot 10^{-5}$ & $2.5\cdot 10^{-5}$  \\
\hline
$Ra=10^7-\varepsilon=0.05$  & $600$  &  $800$ & $5.5\cdot 10^{-5}$ & $5.5\cdot 10^{-5}$  \\
$Ra=10^7-\varepsilon=0.10$  & $600$  &  $800$ & $5.5\cdot 10^{-5}$ & $5.5\cdot 10^{-5}$  \\
$Ra=10^7-\varepsilon=0.15$  & $600$  &  $800$ & $5.5\cdot 10^{-5}$ & $5.5\cdot 10^{-5}$  \\
$Ra=10^7-\varepsilon=0.20$  & $600$  &  $800$ & $5.5\cdot 10^{-5}$ & $5.5\cdot 10^{-5}$  \\
\hline
$Ra=10^8-\varepsilon=0.05$  & $1000$ &  $1200$ & $5.5\cdot 10^{-5}$ & $5.5\cdot 10^{-5}$  \\
$Ra=10^8-\varepsilon=0.10$  & $1000$ &  $1200$ & $5.5\cdot 10^{-5}$ & $5.5\cdot 10^{-5}$  \\
$Ra=10^8-\varepsilon=0.15$  & $1000$ &  $1200$ & $5.5\cdot 10^{-5}$ & $5.5\cdot 10^{-5}$  \\
$Ra=10^8-\varepsilon=0.20$  & $1000$ &  $1200$ & $5.5\cdot 10^{-5}$ & $5.5\cdot 10^{-5}$  \\
\hline
\end{tabular}
\caption{Time window for statistical sampling ($T_{avg}$) and the fixed time step $\Delta t_{avg}$ employed to collect the statistics, both for the cases with (WT) and without  (WO) evaporation. Time is reported in units of free-fall time $\widehat{t}_{ff}$. \nsc{The cases where the gas density and the gas-liquid diffusion coefficient are the only variable properties are conducted at $Ra=10^6$, $10^7$ and $10^8$, and for $\varepsilon=0.05$, $0.10$, $0.15$ and $0.20$. The cases where all the gas thermophysical properties are varied are conducted at $Ra=10^6$ and $10^8$, and for $\varepsilon=0.05$, $0.10$, $0.15$ and $0.20$.}}
\label{tab:set_up}
\end{table}
The computational domain is a two-dimensional cavity with aspect ratio $\mathcal{A}=2$, periodic in the horizontal direction and with two walls at the bottom and top. Here, a no-slip condition is applied for the velocity, a Dirichlet condition for the temperature, while zero-flux is imposed on the remaining quantities. In all the cases, we employ the same uniform grid with $N_x=1024$ and $N_z=512$ along the periodic and the wall-normal directions. This resolution fulfills the two requirements proposed in~\cite{shishkina2010boundary}. Taking as a reference the case at $Ra=10^8$ and $\varepsilon=0.20$ (the most demanding among our cases in terms of grid resolution), we perform two checks summarised below. \par

First, we ensure that the grid size $\widehat{\Delta}=\widehat{l}_z/N_z$ is smaller than the Kolmogorov length-scale,
\begin{equation}
  \widehat{\Delta}_b \leq \pi\widehat{\eta} \approx \pi\widehat{l}_z\sqrt[4]{\dfrac{Pr^2}{RaNu}}\mathrm{.}
  \label{eqn:delta_eta}
\end{equation}
Using the a-posteriori estimation of the Nusselt number $Nu\approx 40$ at $Ra=10^8$ and $\varepsilon=0.20$, we get $\widehat{\Delta}/(\pi\widehat{\eta})\approx 0.20$, thus the requirement is well met. When $\mu$, $\rho$ and $c_p$ are not uniform, $Pr_g$ slightly increase above unity up to $1.53$. Therefore, even though the Bachelor scale is smaller than the Kolmogorov scale, the limited $Pr$ makes the employed resolution suitable for our configuration. \par
\begin{figure*}
  \centering
  \includegraphics[width=6.5 cm, height=5.128 cm]{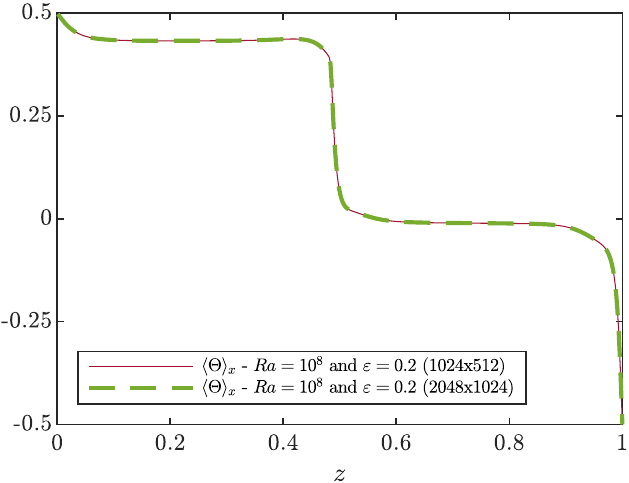}
  \hspace{0.2 cm}
  \includegraphics[width=6.5 cm, height=5.000 cm]{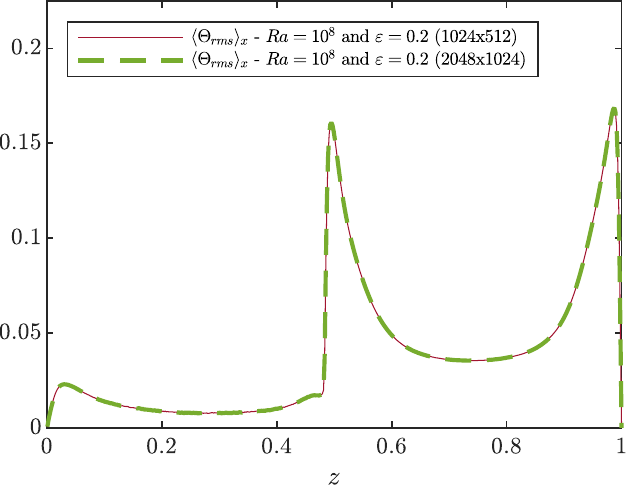}
  \put(-380,130){\small(\textit{a})}
  \put(-195,130){\small(\textit{b})}
  \caption{Grid convergence studies for the case at $Ra=10^8$ and $\varepsilon=0.20$: (\textit{a}) mean vertical profile, (\textit{b}) r.m.s of temperature using $1024\times 512$ and $2048\times 1024$ grid points.}
  \label{fig:finer_grid}
\end{figure*}
Next, we estimate the minimum number of grid points required to fully resolve the thermal and hydrodynamic boundary layers of the gas phase $N_{gp}$, which represents the most stringent condition. Following once more~\cite{shishkina2010boundary}, this requirement reads as,
\begin{equation}
  N_{gp} = \sqrt{2}aNu_g^{1/2}Pr^{0.3215+0.011\log(Pr)}\mathrm{,}
  \label{eqn:delta_ngp}
\end{equation}
with $a=0.482$ in~\eqref{eqn:delta_ngp} and the Nusselt number on the gas side $Nu_g$ estimated as
\begin{equation}
  Nu_g = \dfrac{\widehat{Q}_g\widehat{h}_g}{\widehat{k}_{g,r}(\widehat{T}_\Gamma-\widehat{T}_t)}=Nu\dfrac{\alpha_0f_{g,h}}{(\Theta_\Gamma+1/2)}\mathrm{.}
\end{equation}
Taking $\alpha_0=0.5$, $f_{g,h}=1.05$ and $\Theta_\Gamma=0.36$ in equation~\eqref{eqn:delta_ngp}, $N_{gp}$ is equal to $5$, significantly less than the value of $10$ employed in the current set-up. Finally, given the spatial variation of $\rho_g$ and $D_{lg}$ with the state variables, the local $Sc$ may differ from unity. Taking once more the case at $Ra=10^8$ and $\varepsilon=0.2$, the mean gas density and the vapour diffusion coefficient become $1.8$ and $0.9$ times the corresponding reference values, corresponding to an effective $Sc_{eff}\approx 0.6$. Since $Sc_{eff}<Pr$, the velocity and thermal fields impose the stricter resolution requirement. \par
We have performed a mesh convergence study for the case at $Ra=10^8$ and $\varepsilon=0.2$ by doubling the grid in both directions (i.e.\ $2048\times 1024$) and comparing the result with the corresponding coarser simulation. As shown in figure~\ref{fig:finer_grid}, the chosen resolution ($1024\times 512$) guarantees excellent spatial convergence for first and second-order temperature statistics, confirming once more that it can be considered as adequate for the current study.
%
%
\section{Results and discussions}\label{sec:assess}
\subsection{Temporal evolution of $Nu$ and $\Theta_\Gamma$}
\begin{figure*}
  \centering
  \includegraphics[width=6.5 cm, height=5.0 cm]{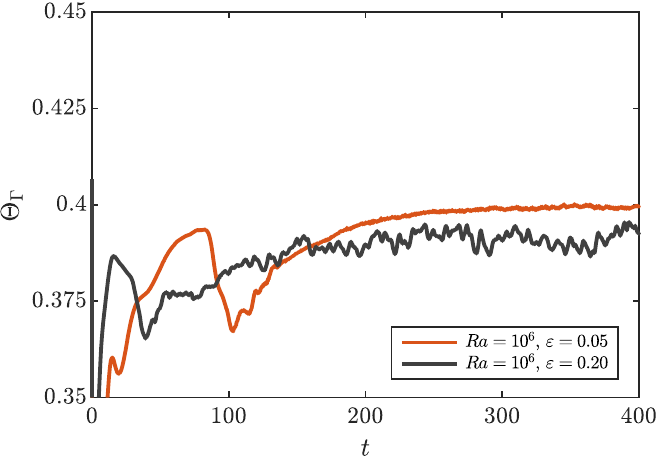}
  \includegraphics[width=6.5 cm, height=5.0 cm]{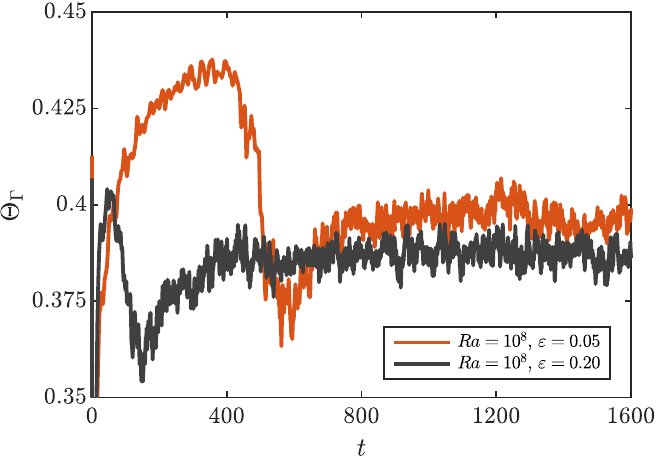}
  \put(-380,130){\small(\textit{a})}
  \put(-190,130){\small(\textit{b})}
  \caption{Temporal evolution of the interface temperature, $\Theta_\Gamma$, for (\textit{a}) $Ra=10^6$ and (\textit{b}) $10^8$, considering $\varepsilon=0.05$ and $0.20$. The instant $t_{eq}=0$ refers to time evaporation at the gas-liquid interface is activated.}
  \label{fig:tmp_time}
\end{figure*}

\nsc{
All the cases are first simulated without evaporation until a statistically stationary equilibrium is reached. Once this condition is met, evaporation is activated at the gas-liquid interface. During the phase-change process, the liquid height is reduced, and the gas region changes its mean temperature, composition and pressure due to increased vapour content. Moreover, phase-change introduces another heat transfer component, i.e.\ latent heat, which is responsible for the mismatch between the Nusselt number values measured at the top and the bottom of the cell, $Nu_t$ and $Nu_b$, during the transient phase. This condition is clearly displayed in figure~\ref{fig:nusselt_time} for $Ra=10^6$ and $10^8$, and $\varepsilon=0.05$, and $0.20$. In particular, $Nu_b$ rapidly increases since evaporation is an endothermic process, while $Nu_t$ rapidly reduces since less heat is transported from the interface to the upper wall. Note that, irrespective of $Ra$, the transient between the two statistical equilibria is faster for larger values of $\varepsilon$ since more vapour is released in the gas layer. Eventually, the gas layer saturates with a balance of evaporation and condensation at the gas-liquid interface, and a new statistical equilibrium is achieved. In this new condition, the time-averaged mass-flux is zero, i.e.\
\begin{equation}
  \int_{t_{eq}}^{t_{eq}+T}\left(\int_\Gamma \dot{m}_\Gamma dS\right)dt=0
  \label{eqn:mean_mfx}
\end{equation}
where $t_{eq}$ is the physical time at which a statistical equilibrium is reached, and $T$ is the time window employed for statistical sampling. Accordingly, the latent heat exchanged at the interface is statically zero, only sensible heat is exchanged, and, thus, the Nusselt number values measured at the top and bottom walls converge to the same statistical mean.
\begin{figure*}
  \centering
  \includegraphics[width=6.5 cm, height=5.0 cm]{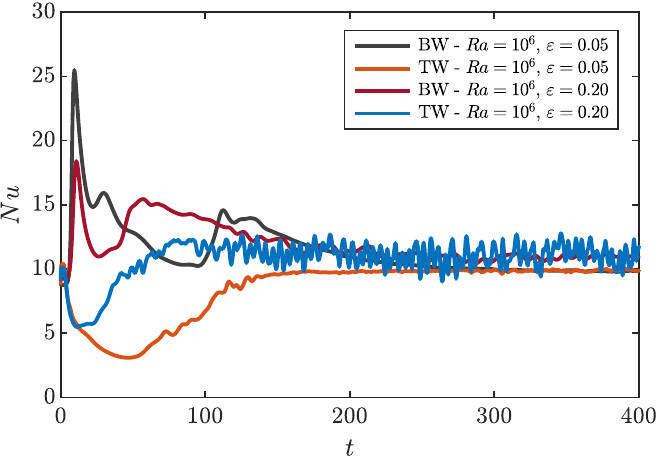}
  \includegraphics[width=6.5 cm, height=5.0 cm]{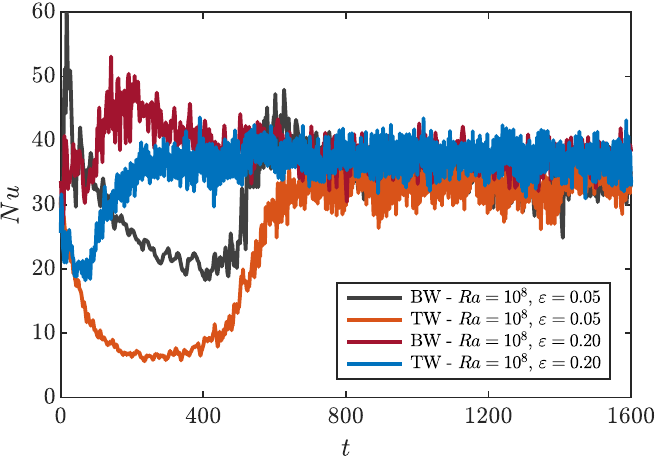}
  \put(-380,130){\small(\textit{a})}
  \put(-190,130){\small(\textit{b})}
  \caption{Temporal evolution of the Nusselt number, $Nu$, for (\textit{a}) $Ra=10^6$ and (\textit{b}) $10^8$, considering $\varepsilon=0.05$ and $0.20$, measured at the bottom wall (BW) and top wall (TW). The time instant $t_{eq}=0$ refers to the instant when evaporation at the gas-liquid interface is activated.}
  \label{fig:nusselt_time}
\end{figure*}
This final condition is also displayed in figure~\ref{fig:nusselt_time} for $Ra=10^6$ and $10^8$, and $\varepsilon=0.05$, and $0.20$. Higher $\varepsilon$ leads to a larger $\overline{Y}_l^v$ and, thus, to an enhancement of the heat transfer in the cell. Moreover, large values of  $\overline{Y}_l^v$ increase the fluctuations of the Nusselt number around its mean value, an effect more pronounced at lower $Ra$. \par
Figure~\ref{fig:tmp_time} displays the time evolution of the interface temperature $\Theta_\Gamma$ for the same cases. When evaporation is active, the interface cools and $\Theta_\Gamma$ suddenly drops due to the latent heat. The final statistically stationary condition is reached when condition~(\ref{eqn:mean_mfx}) is met. Similarly to the temporal behavior of $Nu$, larger values of $\varepsilon$ cause more significant fluctuations of the interface temperature $\Theta_\Gamma$ and reduce the transient before the saturation condition is met. }
\subsection{Validation of the model}
\nsc{
The model described in \S~\ref{sec:scaling} is here validated against two-dimensional interface-resolved direct numerical simulations (DNS) of the evaporating system. We first study the variation of $\rho$, $c_p$, $\mu$ and $k$, together with the molar mass $\overline{M}_m$, with the temperature differential $\varepsilon$. The results, displayed in figure~\ref{fig:var_eos}, are computed by combining the analytical model described in \S~\ref{sec:theta_g} together with the equations of state reported in the appendix~\ref{sec:eos}. For this reason, a non-negligible dependence on the scaling exponent $\gamma$ is observed, especially for the gas density and the molar mass.
%
%
\begin{figure*}
  \centering
  \includegraphics[width=6.5 cm, height=5.000 cm]{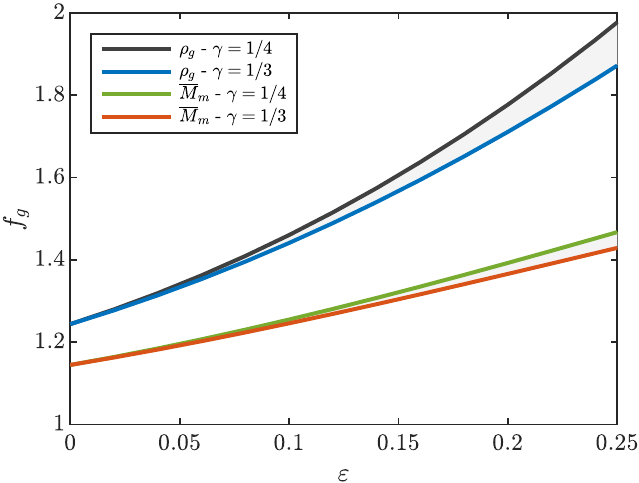}
  \includegraphics[width=6.5 cm, height=4.945 cm]{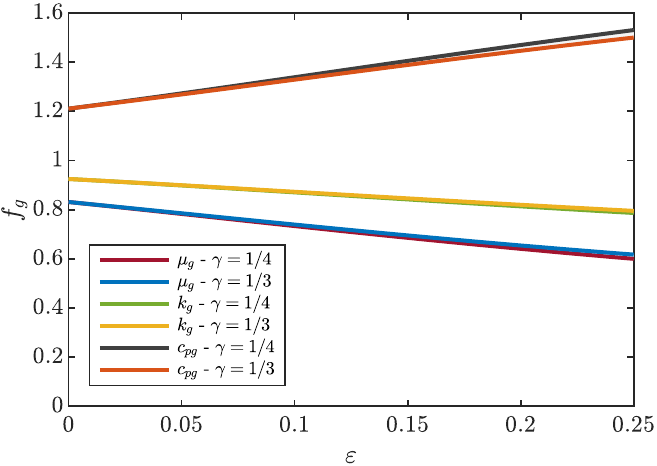}
  \put(-375,130){\small(\textit{a})}
  \put(-190,130){\small(\textit{b})}
  \caption{(\textit{a}) Variation of the normalized mean density $f_{g,\rho}$ and molar mass $\overline{M}_m$ as a function of $\varepsilon$. (\textit{b}) Variation of the normalized mean dynamic viscosity $f_{g,\mu}$, thermal conductivity $f_{g,k}$ and specific heat capacity $f_{g,cp}$. $f_{g,i}$ with $i=\rho$, $\mu$, $k$ and $c_p$ are computed using the definition, i.e.\ equation~\eqref{eqn:xi_g}, and the equations of state detailed in appendix~\ref{sec:eos}.}
  \label{fig:var_eos}
\end{figure*}
The current set-up considers a mixture of air and light hydrocarbon with molar mass ratio $\lambda_M$ larger than 1. Therefore $\overline{M}_m$ increases with $\varepsilon$. Also $f_{g,\rho}$, computed with equation~\eqref{eqn:rhge} increases with the temperature differential $\varepsilon$, but, as shown in figure~\ref{fig:var_eos}, the variation of $f_{g,\rho}$ is dominant over $\overline{M}_m$. Figure~\ref{fig:var_eos} displays the dependence of dynamic viscosity, thermal conductivity, and specific heat capacity with $\varepsilon$. Since the specific heat capacity of the gas, $c_p$, increases with the mean vapour content, its normalized variation, $f_{g,cp}>1$, and increases with $\varepsilon$. The opposite occurs for $\mu_g$ and $k_g$; therefore, their normalized variations, $f_{g,\mu}$ and $f_{g,k}$ are lower than unity with a decreasing trend with $\varepsilon$. \par
To better disentangle the effects of evaporation in the Rayleigh-Benard cell, we start the discussion by considering a subset of the general model, where the gas density and the liquid-gas diffusion coefficient are the only thermophysical properties that vary. This simplification allows us to isolate the effect of the density variations and omit, without further approximations, the Prandtl number dependence in the scaling $Nu_g\sim Ra_g^\gamma$, since the Prandtl number is not a function of $\rho_g$. In this simplified setting, $f_{g,\mu}=f_{g,cp}=f_{g,k}=1$ and equation~\eqref{eqn:theta_g} for the interface temperature $\Theta_{\Gamma}$ becomes
\begin{equation}
  \Theta_{\Gamma} = -\dfrac{1}{2}+\left(1+\left(\dfrac{\alpha_0}{1-\alpha_0}\dfrac{f_{l,h}}{f_{g,h}}\right)^{\dfrac{1-3\gamma}{1+\gamma}}\left(\dfrac{f_{g,\rho}^2\lambda_{\mu}}{\lambda_{\rho}^2\lambda_{cp}\lambda_{\beta}}\right)^{\dfrac{\gamma}{1+\gamma}}\left(\dfrac{1}{\lambda_k}\right)^{\dfrac{1-\gamma}{1+\gamma}}\right)^{-1}\mathrm{.}
  \label{eqn:theta_g1_yes_eva}
\end{equation}
Similarly, the Nusselt number ratio $Nu^e/Nu$ reads
\begin{equation}
  \dfrac{Nu^e}{Nu} = \left(\dfrac{1+2\varepsilon\Theta_c}{1+2\varepsilon\Theta_c^e}\right)^{\gamma}\left(\dfrac{\Theta_{\Gamma}^e+1/2}{\Theta_{\Gamma}+1/2}\right)^{1+\gamma}\dfrac{f_{g,\rho}^{2\gamma,e}}{f_{g,h}^{1-3\gamma,e}}\mathrm{.}
  \label{eqn:nuE_nu}
\end{equation}
Figure~\ref{fig:theta_g_NuE_o_n} compares the analytical prediction of $\Theta_{\Gamma}$ and $Nu^e/Nu$ against the values extracted from the DNS. For both quantities, the predicted results agree very well with the theory, for a choice of scaling exponent $1/4\leq\gamma\leq 1/3$, as suggested by the GL theory (gray region). \par

The data also reveal different aspects on the role of the temperature differential $\varepsilon$ on $\Theta_\Gamma$. First, $\Theta_{\Gamma}$ shows no or little dependence on $\varepsilon$ in a dry environment (i.e.\ no evaporation). This suggests that strong density variations have a negligible impact on $\Theta_\Gamma$ and $Nu$. Second, when evaporation is active, the gas mixture has higher density and $\Theta_{\Gamma}$ decreases with respect to the case without evaporation, exhibiting a stronger-than-linear dependence on $\varepsilon$. Based on the proposed model, the increase of the gas density is expected to occur regardless of the value of $\lambda_M$ (equation~\eqref{eqn:rhge}). Nonetheless, $\lambda_M$ affects the statistically-steady value of $p_{th}$. For $\lambda_M<1$, typical of mixtures of water vapour and inert gas, $\overline{M}_{m}<1$ and, therefore, variations of density and molar mass, $f_{g,\rho}$ and $\overline{M}_{m}$, contribute to the increase of $p_{th}$ (equation.~\eqref{eqn:pth_ratio}). The opposite occurs for $\lambda_M>1$ as in mixtures of hydrocarbons and inert gases, where there is a competitive effect between $f_{g,\rho}$ and $\overline{M}_{m}$, with the former dominating on the latter in the current set-up (see figure~\ref{fig:var_eos}). Finally, note that the increase in the gas density with $\varepsilon$, i.e.\ $f_{g,\rho}>1$, compensates the drop in the interface temperature and, thus contributes to an enhancement of the global heat transfer, i.e.\ $Nu^e/Nu>1$.

\begin{figure*}
  \centering
  \includegraphics[width=6.5 cm, height=5.000 cm]{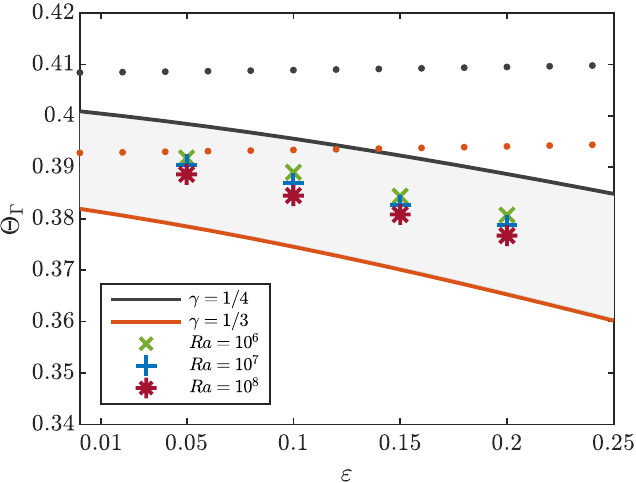}
  \includegraphics[width=6.5 cm, height=4.945 cm]{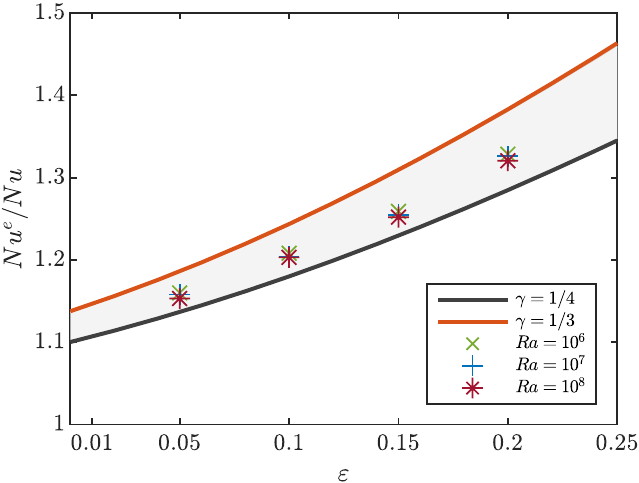}
  \put(-375,130){\small(\textit{a})}
  \put(-190,130){\small(\textit{b})}
  \caption{Comparison between the analytical predictions (with $\gamma\in[1/4,1/3]$) and the numerical simulations for (\textit{a}) the interface temperature $\Theta_{\Gamma}$ and (\textit{b}) the ratio $Nu^e/Nu$ as a function $\varepsilon$ for different values of $Ra$, when only the gas density is varied. Note that in (\textit{a}), the dotted lines correspond to the prediction of $\Theta_\Gamma$ without evaporation.}
  \label{fig:theta_g_NuE_o_n}
\end{figure*}
%
%
We now consider the general case when all the gas thermophysical properties are varied with temperature and composition. Note that the dependence on the thermodynamic pressure is assumed only for the gas density. Based on the parameters and the equations of state we employ for $\mu$, $c_p$, and $k$ (reported in the appendix~\ref{sec:eos}), the temperature dependence is weaker than the effect of the composition. The variation of $\overline{Y}_l^v$ drives the change of $c_p$, $\mu$, and $k$ compared to the dry case. More specifically and as anticipated in figure~\ref{fig:var_eos}, $c_{pg}$ increases with $\varepsilon$, i.e.\ $f_{g,cp}>1$, while $k$ and $\mu$ decreases with $\varepsilon$, i.e.\ $f_{g,\mu}<1$ and $f_{g,k}<1$. Note that from equation~\eqref{eqn:theta_g}, $f_{g,cp}>1$ and $f_{g,\mu}<1$ promote a reduction of the interface temperature in the case of evaporation. Conversely, $f_{g,k}<1$, promotes an increase in $\Theta_\Gamma$. For the current choice of parameters, the changes in viscosity and heat capacity have a stronger effect compared to the thermal conductivity. This factor leads to a decrease of $\Theta_\Gamma$ with $\varepsilon$. Figure~\ref{fig:theta_g_NuE_o_n} displays the analytical predictions for these two quantities and the results from direct numerical simulations conducted for $Ra=10^6$ and $10^8$, and $\varepsilon=0.05$, $0.10$, $0.15$ and $0.20$. As for the other cases, the scaling exponent $\gamma$ is chosen in the interval $[1/4-1/3]$. Compared to the simplified setting where the gas density is the only variable thermophysical property, $\Theta_\Gamma$ exhibits slightly lower values (see figure~\ref{fig:theta_g_NuE_o_n}\textit{(a)}). This behavior is attributed to the larger sensitivity of the interface temperature to $c_p$ and $\mu$ than to variations of $k$. By comparing figure~\ref{fig:theta_g_NuE_o_n}\textit{(b)} with figure~\ref{fig:theta_g_NuE_o_n_all}\textit{(b)}, we note that accounting for the variability of $\mu$, $c_p$, and $k$ leads to a more significant increase of the heat transfer, $Nu^e/Nu$. As shown in equation~\eqref{eqn:qte_qt_1}, the increase of $Nu^e/Nu$ with $\varepsilon$ is driven by the decrease in the heat capacity and viscosity ratio. This increase is only partially compensated by the increase in the thermal conductivity ratio. \par
%
%
\begin{figure*}
  \centering
  \includegraphics[width=6.5 cm, height=5.000 cm]{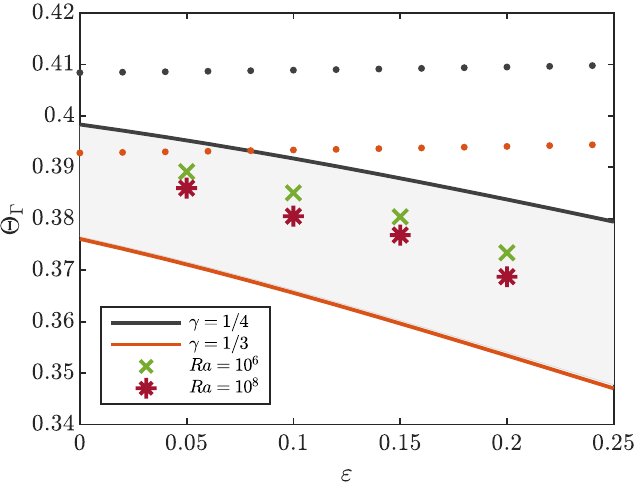}
  \includegraphics[width=6.5 cm, height=4.945 cm]{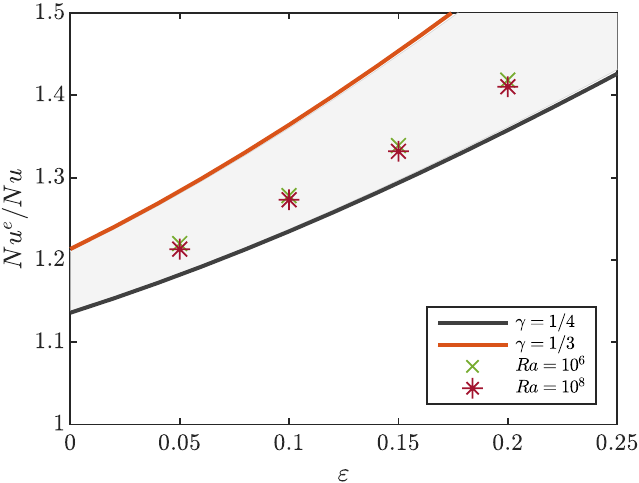}
  \put(-375,130){\small(\textit{a})}
  \put(-190,130){\small(\textit{b})}
  \caption{Comparison between the analytical predictions (with $\gamma\in[1/4,1/3]$) and the numerical simulations for (\textit{a}) the interface temperature $\Theta_{\Gamma}$ and (\textit{b}) the ratio $Nu^e/Nu$ as a function $\varepsilon$ for different values of $Ra$, when all the gas thermophysical properties are varied. Note that in (\textit{a}), the dotted lines correspond to the prediction of $\Theta_\Gamma$ without evaporation.}
  \label{fig:theta_g_NuE_o_n_all}
\end{figure*}
It is worth noticing that in all the cases and irrespective of which gas thermophysical properties vary, the solution does not converge to the case without evaporation for  $\varepsilon\rightarrow 0$, i.e.\ $\Theta_\Gamma^e/\Theta_\Gamma \nrightarrow 1$ and $Nu^e/Nu \nrightarrow 1$ for $\varepsilon\rightarrow 0$. This can be explained by considering the Span-Wagner equation of state as in equation~\eqref{eqn:rault_sw_par} and, in particular, the parameter $\eta_{sw}=1-\Pi_T(1+2\varepsilon)$. When $\varepsilon$ is reduced, $\eta_{sw}$ approaches $1-\Pi_T$ and, therefore, some vapour still reaches in the gas region, i.e.\ $\overline{Y}_{l,\Gamma}^v>0$. The presence of vapour changes the local composition, modify the bulk thermophysical properties and affects both $\Theta_\Gamma$ and $Nu^e$ even for $\varepsilon\rightarrow 0$.
}

%
%
%
\subsection{Assessment of the hypotheses}
\begin{figure*}
  \centering
  \includegraphics[width=6.5 cm, height=5.050 cm]{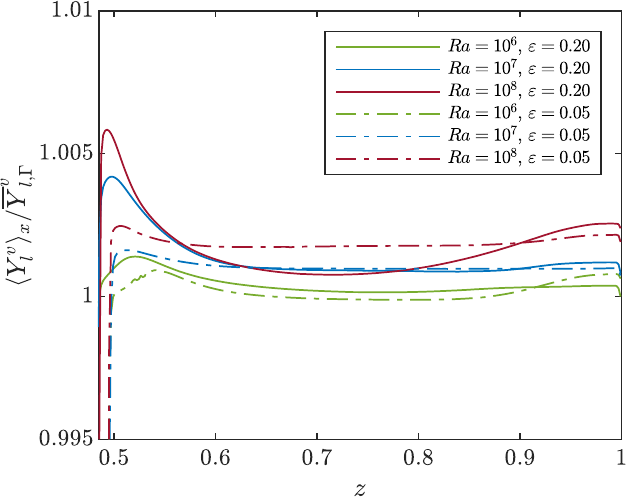}
  \caption{Vertical distribution of the vapour mass fraction $\langle Y_{l}^v\rangle_x$ (normalized by $\overline{Y}_{l,\Gamma}^v$) for $Ra=10^6$, $10^7$ and $10^8$, and $\varepsilon=0.05$ (dot-dashed lines) and $\varepsilon=0.2$ (solid lines).}
  \label{fig:tmp_sca_z}
\end{figure*}

\nsc{
We finally assess the validity of the assumptions invoked at the beginning. With regards to the first assumption, it is reasonable to neglect the variation of the liquid thermophysical properties. This assumption can be easily relaxed by using the mathematical framework proposed here to include appropriate equations of the state for the liquid thermophysical properties. \par
Moving to the second assumption, previous studies in single and multiphase thermal convection have already proven that the GL theory accurately predicts the Nusselt number in the case of NOB effects~\citep{weiss2018bulk} and  two phases~\citep{liu2021heat}. In this work, we have employed a simplified scaling of the form $Nu=A Ra^\gamma Pr^m$, rather than the complete GL theory. Since the simplified scaling is an explicit relation for $Nu$ as a function of $(Ra, Pr)$, we could also derive explicit laws for $\Theta_\Gamma$ and $Nu^e/Nu$ as shown in \S~\ref{sec:scaling}. This would not be possible if adopting the complete GL theory, which provides an implicit relation for $(Nu,Re)$ as a function of $(Ra,Pr)$~\citep{grossmann2000scaling,grossmann2001thermal}. Even though it is possible to extend the present model with the complete GL theory without conceptual modifications, the simplified scaling is still a valid approximation for the present set-up since the liquid Rayleigh and Prandtl number, $Ra_l$ and $Pr_l$, are similar to those in the gas phase, $Ra_g$ and $Pr_g$. It is worth emphasising  that the GL theory ceases to be valid when the interface breaks and significant topological changes occur, as shown in~\cite{liu2021heat}.

To confirm the validity of the last assumption, we employ the results of the DNS. Figure~\ref{fig:tmp_sca_z}
displays the mean vertical profile of $Y_{l}^v$, normalized by $\overline{Y}_{l,\Gamma}^v$ for the different flow configurations under investigation. In all cases, we observe a small positive deviation from the interface values, less than 1 $\%$ for the highest $\varepsilon$, which confirms the validity of approximating $Y_{l}^v$ with $\overline{Y}_{l,\Gamma}^v$. \par 
Finally, it is worth mentioning that despite the overall model is assessed with two-dimensional simulations, we believe its validity is going to be confirmed also in a three-dimensional configurations without any apparent modification. As discussed in~\cite{van2013comparison}, the simplified scaling $Nu\sim Ra^\gamma$ is valid both in two and three dimensions and none of the three assumptions set restrictions on the dimensionality of the problem.}
%
%
%
%
%
\section{Conclusions}\label{sec:concl}
\nsc{
We propose a model for the analytical estimation of the interface temperature $\Theta_{\Gamma}$ and the heat transfer modulation, quantified by the Nusselt number, for an evaporating two-layer Rayleigh-B\'enard configuration at a statistically stationary state. The model is based on three assumptions: (i) the Oberbeck-Boussinesq approximation can be applied to the liquid phase, while the gas thermophysical properties are generic functions of the thermodynamic pressure, local temperature and vapour composition, (ii) the Grossmann-Lohse theory for thermal convection can be applied to the liquid and gas layers separately, (iii) the vapour content in the gas can be taken as the mean value at the gas-liquid interface. The model provides a quantitative prediction of the ratio $Nu^e/Nu$, enabling us to predict the global heat transfer of the evaporating system once the value for the same system in dry conditions, i.e.\ without phase change, is known. 
We validate the analytical predictions using direct numerical simulations in the low-Mach number regime in a parameter space defined by $10^6\leq Ra\leq 10^8$ and $0.05\leq \varepsilon\leq 0.20$. 
Simulations are performed in two settings: (i) assuming the gas density and liquid-diffusion coefficient are the only variable property, (ii) in the general case where all the gas thermophysical properties depend on the state variables. Irrespective of the setting, a very good agreement between the model predictions and the numerical simulations is found by just adopting reasonable values for the scaling exponent of the GL theory, i.e.\ $1/4\leq\gamma\leq 1/3$. Finally, we assess the basic assumptions on which the entire model is built and conclude that they are generally valid unless the interface undergoes large deformation and breakup, which would make the GL theory not valid anymore~\citep{liu2021heat}. We believe that the proposed model and further extensions may find applications where accurate predictions of $\Theta_\Gamma$ and $Nu$ are required and to improve single-phase models of turbulent convection in the presence of evaporation~\citep{schumacher2010buoyancy,hay2020evaporation}.
}
\subsection*{Acknowledgements}
N.S., A.D. and L.B. acknowledge the support from the Swedish Research Council via the multidisciplinary research environment INTERFACE (Pr. 2016-06119). Computer time was provided by the Swedish National Infrastructure for Computing (SNIC) and by the Norwegian research infrastructure services (NRIS, Pr. NN9561K). Dr. Pedro Costa is acknowledged for the useful discussions, and Mr. William Gross is thanked for pointing out some typos in an early version of the manuscript.
\nsc{
\appendix
\section{Equations of state}\label{sec:eos}
We detail the equations of state employed to compute the gas density, specific heat capacity, dynamic viscosity, and thermal conductivity of the gas phase. Note that the gas density is a function of temperature, thermodynamic pressure and vapour composition. The dependence on the thermodynamic pressure is usually negligible for the remaining thermophysical properties~\citep{reid1987properties} and, therefore, it is omitted in the present work.

\subsection{Gas density}\label{subsec:gas_density}
The gas density $\widehat{\rho}_g$ is evaluated with the equation of state for the ideal gas:
\begin{equation}
  \widehat{\rho}_g = \dfrac{\widehat{p}_{th}\widehat{M}_m}{\widehat{R}_u\widehat{T}_g}\mathrm{,}
  \label{eqn:rho_g}
\end{equation} 
where $\widehat{p}_{th}$ is the thermodynamic pressure, $\widehat{R}_u$ is the ideal gas constant, $\widehat{T}_g$ is the gas temperature and $\widehat{M}_m$ is the molar mass, computed using the harmonic average between the one of the liquid and the one of the gas, i.e.\ $\widehat{M}_m=\left(Y_l^v/\widehat{M}_l+(1-Y_l^v)/\widehat{M}_g\right)^{-1}$. By introducing a reference thermodynamic pressure $\widehat{p}_{th,r}$, a reference temperature difference $\widehat{\Delta T}=2\varepsilon\widehat{T}_r$, reference molar mass $\widehat{M}_{g,r}=\widehat{M}_{g}$ and using $\Theta_g=(\widehat{T}_g-\widehat{T}_r)/\widehat{\Delta T}$, equation~\eqref{eqn:rho_g} can be written in dimensionless form as
\begin{equation}
  \rho_g = \dfrac{p_{th}M_m}{1+2\varepsilon\Theta_g}\mathrm{.}
  \label{eqn:rho_g_dim}
\end{equation}
Note that in equation~\eqref{eqn:rho_g_dim}  the reference density $\widehat{\rho}_{g,r} = \widehat{p}_{th,r}\widehat{M}_g/(\widehat{R}_u\widehat{T}_r)$. 

\subsection{Specific heat capacity}\label{subsec:heat_capacity}
The specific heat capacity $\widehat{c}_{pg}$ of the gas phase is computed as linear combination between the one of the vapour, $\widehat{c}_{p,v}$, supposed to be independent of temperature, and the one of the dry gas, $\widehat{c}_{p,dg}$ as
\begin{equation}
  \widehat{c}_{pg} = \widehat{c}_{p,v}Y_l^v + \widehat{c}_{pg,d}(1-Y_l^v)\mathrm{.}
  \label{eqn:cp_g}
\end{equation} 
The gas heat capacity in dry conditions, $\widehat{c}_{pg,d}$, is a function of temperature and it is computed as
\begin{equation}
  \widehat{c}_{pg,d} = \widehat{C}_1+\widehat{C}_2((\widehat{C}_3/\widehat{T}_g)/(\sinh(\widehat{C}_3/\widehat{T}_g)))^2 + \widehat{C}_4((\widehat{C}_5/\widehat{T}_g)/(\sinh(\widehat{C}_5/\widehat{T}_g)))^2\mathrm{,}
  \label{eqn:cp_dg}
\end{equation} 
where $\widehat{C}_{i=1,5}$ are semi-empirical constants for the dry gas taken equal to $\widehat{C}_{i=1,5}$=[$4.13\cdot 10^{4}$ J/(kmol$\cdot$ K), $1.34\cdot 10^{4}$ J/(kmol$\cdot$ K), $3012$ K, $1.08\cdot 10^{4}$ J/(kmol$\cdot$ K), $1484$ K], as suggested in~\citep{reid1987properties}. \par
Equation~\eqref{eqn:cp_g} can be rewritten in dimensionless form using the reference specific heat capacity evaluated at $\widehat{T}_r$ from equation~\eqref{eqn:cp_dg}
\begin{equation}
  c_{pg} = \Pi_{cp}Y_l^v + c_{pg,d}(1-Y_l^v)\mathrm{,}
  \label{eqn:cp_g_dim}
\end{equation} 
where $\Pi_{cp}=\widehat{c}_{p,v}/\widehat{c}_{pg,r}$, in the present work equal to $2.0216$. $c_{pg,d}$ is computed from the dimensionless expression of equation~\eqref{eqn:cp_dg}. By using $\widehat{\Delta T}=2\varepsilon\widehat{T}_r$, equation~\eqref{eqn:cp_dg} can be written in dimensionless form
\begin{equation}
  c_{pg,d} = C_1+C_2((C_3/\Theta_g^*)/(\sinh(C_3/\Theta_g^*)))^2 + C_4((C_5/\Theta_g^*)/(\sinh(C_5/\Theta_g^*)))^2\mathrm{,}
  \label{eqn:cp_dg_dim}
\end{equation} 
where $\Theta_g^*=1+2\varepsilon\Theta_g$ and $C_{i=1,5}=\left [0.6966,0.2259,8.0175,0.1824,3.9502\right]$.

\subsection{Dynamic viscosity}\label{subsec:viscosity}
The dynamic viscosity of the gas phase $\widehat{\mu}_g$  is computed as combination between the one of the vapour, $\widehat{\mu}_v$, supposed to be independent of temperature, and the one of the dry gas, $\widehat{\mu}_{dg}$. Differently from the specific heat capacity, the mixture rule is typically non linear. In the present work, we employ the Wilke-Lee mixture rule as suggested in~\cite{reid1987properties}. This reads as
\begin{equation}
  \widehat{\mu}_g = \dfrac{Y_{l,m}^v\widehat{\mu}_v}{Y_{l,m}^v+(1-Y_{l,m}^v)\phi_{vg}}+\dfrac{(1-Y_{l,m}^v)\widehat{\mu}_{dg}}{Y_{l,m}^v\phi_{gv}+(1-Y_{l,m}^v)}\mathrm{.}
  \label{eqn:mu_g}
\end{equation} 
Equation~\eqref{eqn:mu_g} requires to first evaluate $Y_{l,m}^v$ as
\begin{equation}
  Y_{l,m}^v = \dfrac{Y_l^v}{\widehat{M}_l} + \dfrac{1-Y_l^v}{\widehat{M}_g}\mathrm{.}
  \label{eqn:y_lm}
\end{equation} 
Next, the weighting coefficients $\phi_{vg}$ and $\phi_{gv}$, which are functions of the molar mass of the vapour and the dry gas, are evaluated as
\begin{equation}
  \phi_{vg}=\dfrac{\left(1+\sqrt{\dfrac{\widehat{\mu}_v}{\widehat{\mu}_{g,d}}}\lambda_M^{-1/4}\right)^2}{\sqrt{8\left(1+\lambda_M\right)}}\mathrm{,} \hspace{1 cm}
  \phi_{gv}=\dfrac{\left(1+\sqrt{\dfrac{\widehat{\mu}_{g,d}}{\widehat{\mu}_{v}}}\lambda_M^{1/4}\right)^2}{\sqrt{8\left(1+\lambda_M^{-1}\right)}}\mathrm{.}
  \label{eqn:phi_vi_mu}
\end{equation} 
The gas viscosity can be evaluated with the simplified Sutherland's law
\begin{equation}
  \widehat{\mu}_{g,d} = \widehat{\mu}_{g,r}\left(\dfrac{\widehat{T}_g}{\widehat{T}_r}\right)^{2/3}\mathrm{.}
  \label{eqn:mu_dg}
\end{equation} 
Equation~\eqref{eqn:mu_g} can be rewritten in dimensionless form using the reference viscosity evaluated at $\widehat{T}_r$ from equation~\eqref{eqn:mu_dg}
\begin{equation}
  \mu_g = \dfrac{Y_{l,m}^v\Pi_\mu}{Y_{l,m}^v+(1-Y_{l,m}^v)\phi_{vg}}+\dfrac{(1-Y_{l,m}^v)\mu_{g,d}}{Y_{l,m}^v\phi_{gv}+(1-Y_{l,m}^v)}\mathrm{,}
  \label{eqn:cp_g_dim}
\end{equation} 
where $\Pi_{\mu}=\widehat{\mu}_v/\widehat{\mu}_{g,r}=0.3321$. Moreover, by using $\widehat{\Delta T}=2\varepsilon\widehat{T}_r$, equation~\eqref{eqn:mu_dg} can be written in dimensionless form as
\begin{equation}
  \mu_{g,d} = \left(1+2\varepsilon\Theta_g\right)^{2/3}\mathrm{.}
  \label{eqn:mu_dg_dim}
\end{equation} 

\subsection{Thermal conductivity}\label{subsec:thermal_conductivity}
The thermal conductivity $\widehat{k}_g$ of the gas phase is computed similarly to the gas viscosity, i.e.\ using the non linear Wilke-Lee mixture rule with suitable modifications, i.e.\
\begin{equation}
  \widehat{k}_g = \dfrac{Y_{l,m}^v\widehat{k}_v}{Y_{l,m}^v+(1-Y_{l,m}^v)\phi_{vg}}+\dfrac{(1-Y_{l,m}^v)\widehat{k}_{dg}}{Y_{l,m}^v\phi_{gv}+(1-Y_{l,m}^v)}\mathrm{,}
  \label{eqn:k_g}
\end{equation} 
where $Y_{l,m}^v$ is evaluated with~\eqref{eqn:y_lm}. Next, the weighting coefficients $\phi_{vg}$ and $\phi_{gv}$, which are functions of the molar mass of the vapour and the dry gas, are evaluated using similar expressions as~\eqref{eqn:phi_vi_mu},
\begin{equation}
  \phi_{vg}=\dfrac{\left(1+\sqrt{\dfrac{\widehat{k}_v}{\widehat{k}_{g,d}}}\lambda_M^{-1/4}\right)^2}{\sqrt{8\left(1+\lambda_M\right)}}\mathrm{,} \hspace{1 cm}
  \phi_{gv}=\dfrac{\left(1+\sqrt{\dfrac{\widehat{k}_{g,d}}{\widehat{k}_{v}}}\lambda_M^{1/4}\right)^2}{\sqrt{8\left(1+\lambda_M^{-1}\right)}}\mathrm{.}
  \label{eqn:phi_vi_ka}
\end{equation}
The gas thermal conductivity in dry condition can be evaluated with the following expression
\begin{equation}
  \widehat{k}_{g,d} = \dfrac{\widehat{C}_1\widehat{T}_g^{\widehat{C}_2}}{1+\dfrac{\widehat{C}_3}{\widehat{T}_g}+\dfrac{\widehat{C}_4}{\widehat{T}_g^2}}\mathrm{,}
  \label{eqn:k_dg}
\end{equation} 
where $C_{i=1,4}$ are semi-empirical constants for the dry gas taken equal to $\widehat{C}_{i=1,5}$=[$3.8889\cdot 10^{-4}$ W/(K$\cdot$ m), $0.7786$, $-0.7716$ K, $2121.7$], as suggested in~\citep{reid1987properties}. Equations~\eqref{eqn:k_g} can be rewritten in dimensionless form using the reference thermal conductivity evaluated at $\widehat{T}_r$ from equation~\eqref{eqn:mu_dg}
\begin{equation}
  k_g = \dfrac{Y_{l,m}^v\Pi_k}{Y_{l,m}^v+(1-Y_{l,m}^v)\phi_{vg}}+\dfrac{(1-Y_{l,m}^v)k_{g,d}}{Y_{l,m}^v\phi_{gv}+(1-Y_{l,m}^v)}\mathrm{,}
  \label{eqn:k_g_dim}
\end{equation} 
where $\Pi_{k}=\widehat{k}_v/\widehat{k}_{g,r}=0.5824$. 
$k_{g,d}$ is computed from the dimensionless expression of equation~\eqref{eqn:k_dg}, obtained defying appropriate reference quantities as for the previous thermophysical properties,
\begin{equation}
  k_{g,d} = \dfrac{C_1\Theta_g^{*,C_2}}{1+\dfrac{C_3}{\Theta_g^*}+\dfrac{C_4}{\Theta_g^{*,2}}}\mathrm{,}
  \label{eqn:k_dg_dim}
\end{equation} 
where $C_{i=1,4}=\left [0.3039,0.7786,-0.0021,+5.6476\right]$.
}
\section{Validation}\label{sec:val}
\begin{figure*}
  \centering
  \includegraphics[width=7 cm, height=5.25 cm]{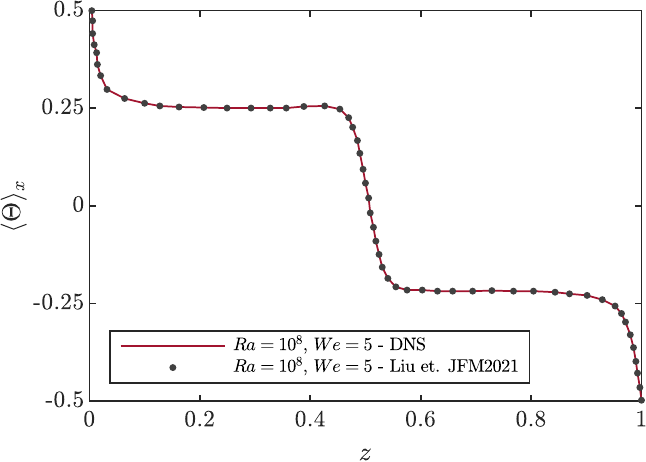}
  \caption{Mean temperature profiles obtained with the numerical code employed in the present study (continuous line) and the results in~\cite{liu2021two} (dotted lines) for a two-dimensional two-fluid RB flow with $Ra=10^8$, $\lambda_\rho=3.33$ and $We=5$.}
  \label{fig:comparison}
\end{figure*}

For completeness, we validate our code using the data from the two-layer Rayleigh-B\'enard convection study in \cite{liu2021two}. Here, the Oberbeck-Boussinesq approximation is assumed for the gas and liquid layers, and phase change is absent. To reproduce this condition, we "switch-off" phase change and solve only equations~\eqref{eqn:h_ind},~\eqref{eqn:mom1} and~\eqref{eqn:tmp1} setting $\varepsilon=0.005$. Moreover, we consider $Ra=10^8$, $\lambda_{\rho}=3.33$ and Weber number $We=5$, while the remaining thermophysical properties ratios $\lambda_\xi=1$. Simulations are conducted in a two-dimensional domain discretized with $N_x=1024$ and $N_z=512$. Figure~\ref{fig:comparison} displays the mean vertical temperature profile. The excellent agreement between our simulations and the reference data in~\cite{liu2021two} validates our numerical algorithm for a two-layer Rayleigh-B\'enard configuration in the OB limit.

\bibliographystyle{jfm}
\bibliography{bibfile.bib}
\end{document}